\documentclass{article}

\usepackage{arxiv}
\usepackage{amssymb}
\usepackage{mathptmx} 
\usepackage{ulem}
\usepackage{fancyhdr}
\usepackage[hyphens]{url}
\usepackage[final]{microtype}
\usepackage[keeplastbox]{flushend}
\usepackage[bookmarks=true,breaklinks=true,letterpaper=true,colorlinks,linkcolor=black,citecolor=blue,urlcolor=black]{hyperref}

\title{Guidelines for Submission to MICRO 2020} 

\usepackage[noadjust]{cite}
\usepackage{mathtools}

\usepackage{color}
\newcommand{\blue}[1]{\textcolor{blue}{#1}} 
\newcommand{\red}[1]{\textcolor{black}{#1}} 

\usepackage{scalerel,stackengine}
\stackMath
\newcommand\reallywidehat[1]{%
\savestack{\tmpbox}{\stretchto{%
  \scaleto{%
    \scalerel*[\widthof{\ensuremath{#1}}]{\kern-.6pt\bigwedge\kern-.6pt}%
    {\rule[-\textheight/2]{1ex}{\textheight}}
  }{\textheight}%
}{0.5ex}}%
\stackon[1pt]{#1}{\tmpbox}%
}
	
\usepackage{algorithm,algorithmicx,algpseudocode}

\usepackage{array}
\usepackage{makecell}

\usepackage{hyperref}
\hypersetup{pdfborderstyle={/S/U/W 1}}

\usepackage{graphicx, subfigure}
\DeclareGraphicsExtensions{.pdf,.png,.jpg}

\usepackage{tikz}
\usetikzlibrary{matrix,calc}
\usetikzlibrary{snakes}

\usepackage{pgfplots}

\usepackage{hhline}

\usepackage{titlecaps}

\usepackage{booktabs,siunitx}
\usepackage{adjustbox}

\usepackage{multirow}

\newcommand{\mn}{MobileNet-V2}
\newcommand{\effnet}{EfficientNet}
\newcommand{\vmn}{MobileNet-V1}
\newcommand{\dw}{depthwise}
\newcommand{\pw}{pointwise}
\newcommand{\nc}{Normal Convolution}

\newcommand{\cn}{convolution}

\newcommand{\gc}{group-convolution}
\newcommand{\htr}{heterogeneous}
\newcommand{\qnet}{\textit{QNet}}

\newcommand{\cmmnt}[1]{}
\begin{document}


\makeatletter
\def\@copyrightspace{\relax}
\makeatother

\title{DeepDive: An Integrative Algorithm/Architecture Co-Design for Deep Separable Convolutional Neural Networks}

\author{Mohammadreza~Baharani,
        Ushma~Sunil,
        Kaustubh~Manohar,
        Steven~Furgurson,
		Hamed~Tabkhi,
\thanks{The authors are with the Electrical and Computer Engineering Department, Energy Production and Infrastructure Center (EPIC), The University of North Carolina-Charlotte, Charlotte,
	NC, 28223 USA  (e-mail: mbaharan@uncc.edu,  ubharuch@uncc.edu,  kmhatre@uncc.edu,  sfurgurs@uncc.edu,  htabkhiv@uncc.edu).}}

\maketitle

\begin{abstract}


Deep Separable Convolutional Neural Networks (DSCNNs) have become the emerging paradigm by offering modular networks with structural sparsity in order to achieve higher accuracy with relatively lower operations and parameters. However, there is a lack of customized architectures that can provide flexible solutions that fit the sparsity of the DSCNNs. This paper introduces DeepDive, which is a fully-functional, vertical co-design framework, for power-efficient implementation of DSCNNs on edge FPGAs. DeepDive's architecture supports crucial heterogeneous Compute Units (CUs) to fully support DSCNNs with various convolutional operators interconnected with structural sparsity. It offers an FPGA-aware training and online quantization combined with modular synthesizable C++ CUs, customized for DSCNNs. The execution results on Xilinx's ZCU102 FPGA board, demonstrate 47.4 and 233.3 FPS/Watt for MobileNet-V2 and a compact version of EfficientNet, respectively, as two state-of-the-art depthwise separable CNNs. These comparisons showcase how DeepDive improves FPS/Watt by 2.2$\times$ and 1.51$\times$ over Jetson Nano high and low power modes, respectively. It also enhances FPS/Watt about 2.27$\times$ \red{and 37.25$\times$ over two other FPGA implementations. The DeepDive output for MobileNetV2 is available at \href{https://github.com/TeCSAR-UNCC/DeepDive}{\ttfamily{\textcolor{blue}{ https://github.com/TeCSAR-UNCC/DeepDive}}}.}





\end{abstract}

\section{Introduction}\label{sec:introduction}
The astonishing growth in deep learning algorithms, particularly, Convolutional Neural Networks (CNNs), has enabled many exciting applications in visual analytics. We have observed a recent shift towards Domain-Specific Architectures (DSA), e.g., Systolic Arrays, CGRAs, Tensor Cores, to cope with the significant computation demand raised by deep learning paradigms \cite{dnnweaver:micro16, angeleye, dnnbuilder, snowflake, sysarrayaccel, Caffeine, DBLP:vta}. These emerging DSAs often transform convolutional operations into dense linear algebraic operations across the channels and kernels. This maximizes parallelism and compute resource utilization, as well as minimizes data movements, by increasing data re-usability. They are typically designed to be a generic, one-size-fits-all architecture that allows hardware reuse between different layer operations. As a result, they execute the target CNN layer-by-layer sequentially. A notable example is the recently introduced Versatile Tensor Accelerator (VTA) \cite{DBLP:vta}, which is an open, generic, and customizable deep learning accelerator with a complete TVM-based compiler stack, targeted for edge FPGAs. \red{Another such accelerator design presented by \cite{mob} introduces a configurable architecture, pipelined, and timing controlled design with fixed hardware solution specially designed for MobileNet.}

\begin{figure}[!t]
    \centering
    \includegraphics[width=.5\textwidth, trim= 15 15 20 20,clip, keepaspectratio]{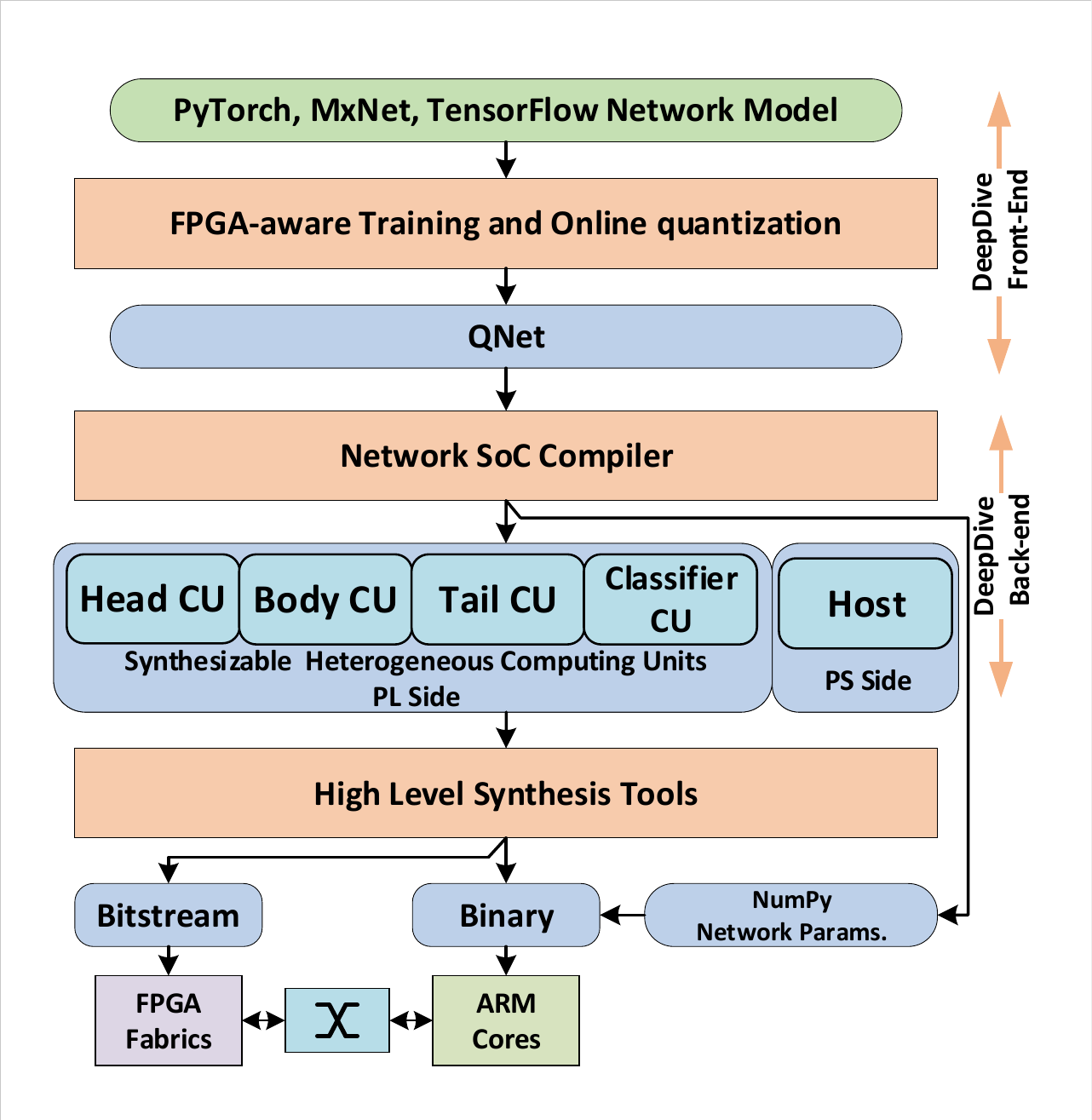}
    \vspace{-7pt}
    \caption{DeepDive integrative design flow.}
     \vspace{-14pt}
    \label{fig:DeepDive_intro}
\end{figure}

Deep Separable CNNs (DSCNNs) \cite{ShuffleNet, ResNext, FD_NET, mobileNetV2, EfficientNet, mobilenet} have emerged as an innovative algorithmic solutions to achieve higher accuracy with relatively lower parameters and operations. State-of-the-art separable CNNs, e.g., MobileNet family \cite{mobilenet, mobileNetV2} and EfficientNet \cite{EfficientNet}, offer modular networks with structural sparsity over various convolutional operators --- group, \dw, and \pw~\cn. DSCNNs often result in relatively higher computational sparsity, more data-dependent layer-to-layer communication, and less data reuse potential over their predecessor networks, such as ResNet \cite{ResNet} or VGG \cite{VGG}. At the same time, the modular design, combined with the structural sparsity of DSCNNs, allows the designer to systematically trade between algorithmic accuracy, and computational demand, via tunable knobs that vary the sparsity of the network, e.g., varying degree of width multiplication in \mn.

The structural sparsity of DSCNNs makes existing DSAs, e.g., VTA or Tensor Cores, less suitable for efficient execution of DSCNNs, as the current DSAs have been often designed for dense operations with highly regular data access and high data reuse. At the same time, current DSAs are often optimized for a single design point in isolation, which limits their efficiency when running DSCNNs. For instance, they convert sparse convolutions to dense matrices (e.g., \dw~to \gc~transform), which leads to higher computational overhead than the original DSCNNs, while delivering the same accuracy. As an example, VTA had to make a specialized version of MobileNet, which they call MobileNetG, to remove \dw~separable \cn~and make it running efficiently on systolic array implemented on FPGAs. \red{FPGA implementation introduced in \cite{mob} has massive data movements as a result of their configurable data path design which results in high latency. Also such type of fixed architectures adopted in \cite{mob, cnnAccF, mdpiCnn, dpu} makes it difficult to achieve scalability to support modern DSCNNs, e.g. \effnet.}

This paper proposes a fully functional framework called DeepDive for an agile, power-efficient execution of DSCNNs. DeepDive offers a novel architecture for efficient execution of DSCNNs, combined with a vertical algorithm/architecture optimization and synthesis on edge FPGAs. The framework is designed to identify key heterogeneous Compute Units (CUs), to fully support DSCNNs with heterogeneous convolutional operations, such as group, \dw, and \pw~\cn. Fig.~\ref{fig:DeepDive_intro} abstracts DeepDive design flow. At the front-end, DeepDive receives the network description model (e.g., PyTorch), and optimizes the model based on the FPGA-aware training and online quantization. This includes algorithm-specific fusing of batch normalization and convolutional operators, which reduces the computation by \textasciitilde 4\%, and extremely low-bit per-channel-quantization across all separable convolution layers. The output of the front-end will be \qnet, which contains all of the meta-data regarding the FPGA-aware trained as well as quantized network model. At the back-end, DeepDive relies on the recent advances in High-Level Synthesis (HLS) and shifts the optimization abstraction to pre-RTL design. \red{The \textit{Network SoC Compiler} creates a customized memory path and synthesizable model of the entire hardware accelerator for Programmable Logic (PL) based on pre-designed CUs and provided \cn~operators. It also generates the host CPU code~ running on ARM cores located in the Processing System (PS) side of SoC for synchronization and scheduling}. The host code, bundled with a scheduler, enables the DeepDive back-end system to support multiple run-time software stacks such as Pynq and Linux. \red{The key contributions are:
\begin{itemize}
    \item The structure and the flexibility of DeepDive enables an agile framework to support the fast-growing and up-coming DSCNNs. To the best of our knowledge, this work is the first scalable solution with the support of recently introduced \effnet~DSCNN families.
    \item It proposes a novel scalable vertical framework for the execution of DSCNN on FPGAs. The vertical integration and library-based operation mapping enables true comprehensive design space exploration on FPGAs.
\end{itemize}
}
The rest of this article is organized as the following:  Section \ref{sec:HtrgNect} discuss the algorithmic properties of DSCNNs and further motivates DeepDive. Section \ref{sec:DD_F} presents DeepDive's front-end, focusing on FPGA-aware training and online quantization. Section \ref{sec:DD_B} details DeepDive's back-end architecture and design flow. Section \ref{sec:ExpResults} presents DeepDive's execution results on Xilinx's ZCU102 FPGA and comparison against state-of-the-art solutions. Section~\ref{sec:relatedWork} reviews the related work. Finally, Section~\ref{sec:conclusion} concludes this paper.

\section{Algorithmic Principles of Deep Separable CNNs}\label{sec:HtrgNect}
DSCNNs \cite{ShuffleNet, mobileNetV2, EfficientNet, mobilenet} have emerged as a new paradigm to achieve higher accuracy with relatively fewer parameters and operations over the classical CNNs. The efficiency of DSCNNs stems from their structural sparsity, combined with a modular configurable network topology, that can be scaled up or down, depending on desired accuracy and corresponding computational overhead. In this section, we define the basic principles and structural properties of DSCNN. For ease of access, we summarized the symbols that appeared in this paper and their description in Table~\ref{table:symbols_paper}. These symbols will be used throughout this paper.

\begin{table}[!htbp]
	\caption{List of symbols}
		\centering 
		\scalebox{1.0}{
			\begin{tabular}{c c c c c}
				\hline
				\hline
				Item & Parameter \ & Description
				\\[0.1ex]
				\hline              
				1 & $N$ & Input channel size
				\\
				2 & $M$  &Output channel size
				\\
				3 & $K$ & Kernel size
				\\
				4 & $H$ & Height of input feature
				\\
				5 & $W$ & Width of input feature
				\\
				6 & $G$ & Group size
				\\
				7 & $BW$ & Bit-width
                \\
				8 & $\alpha$ & Width multiplier
				\\
				9 & $k$ & Number of classes
				\\
				\hline			
			\end{tabular}
			\label{table:symbols_paper}
		}
\end{table}


\begin{figure}[b]
	\centering
	\subfigure[Normal convolution]{%
		{\includegraphics[width=0.22\textwidth, trim= 20 15 20 20,clip, keepaspectratio]{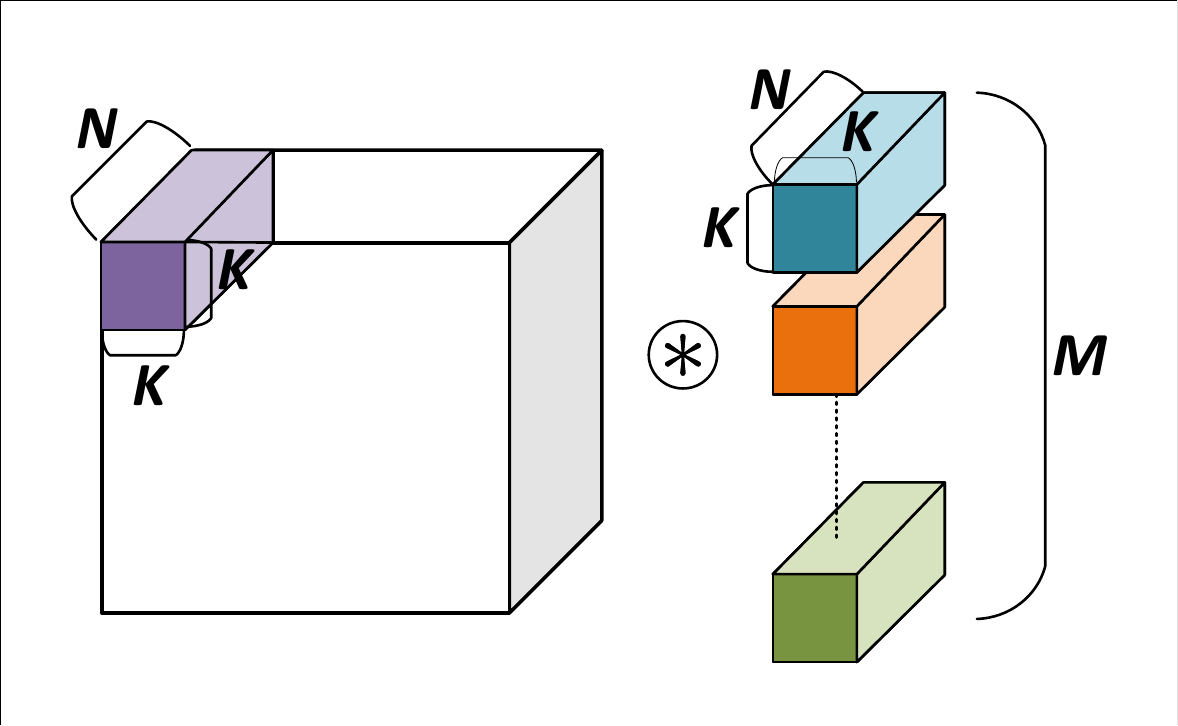}}%
		\label{fig:cnn}%
	}\qquad
	\subfigure[Group convolution]{%
		{\includegraphics[width=0.22\textwidth, trim= 30 15 25 20,clip, keepaspectratio]{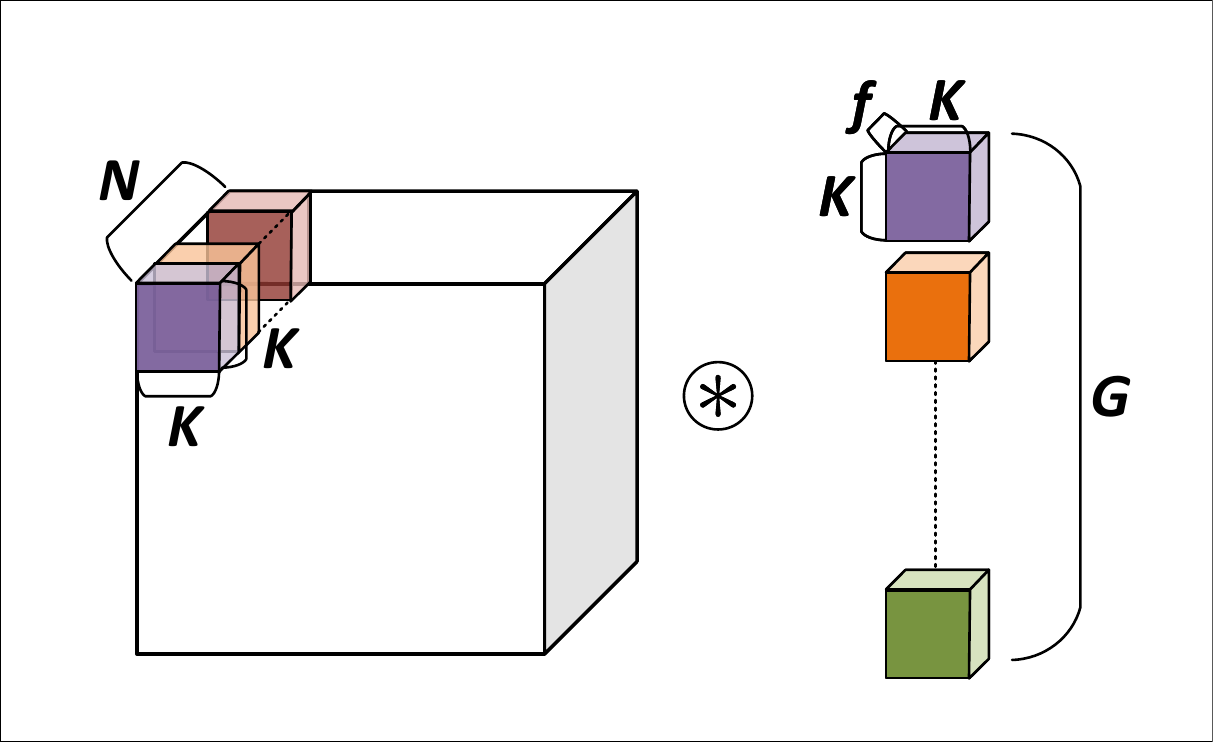}}%
		\label{fig:gc}%
	}\\
	\subfigure[Depthwise convolution]{%
		{\includegraphics[width=0.22\textwidth, trim= 20 15 25 20,clip, keepaspectratio]{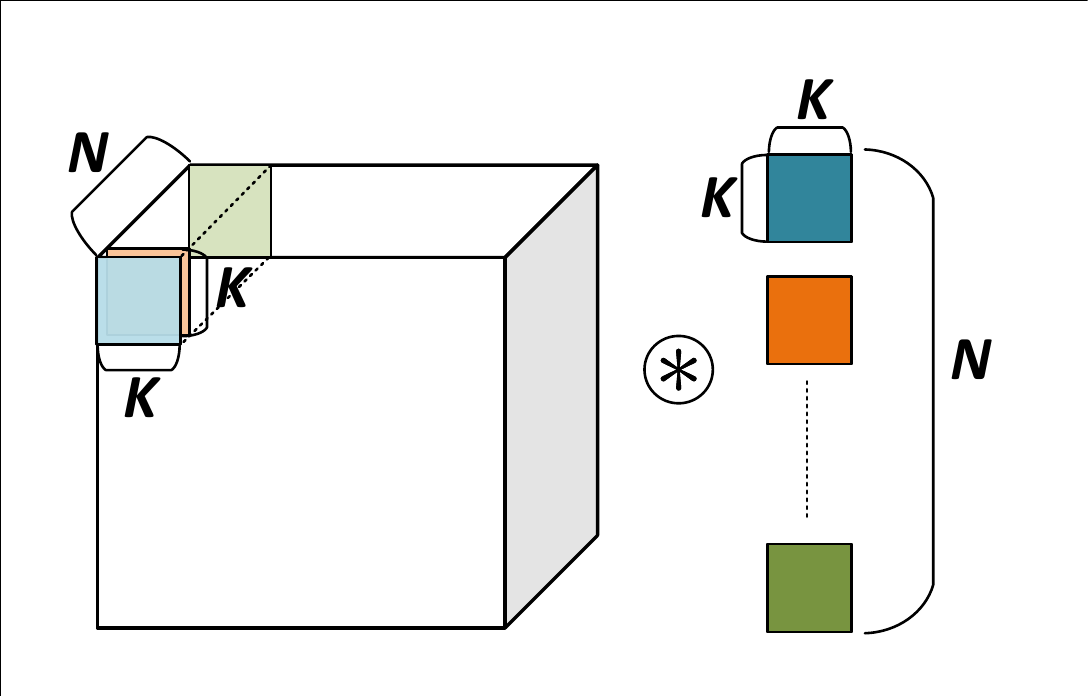}}%
		\label{fig:dw}%
	}\qquad
	\subfigure[Pointwise convolution]{%
	{\includegraphics[width=0.22\textwidth, trim= 20 15 20 20,clip, keepaspectratio]{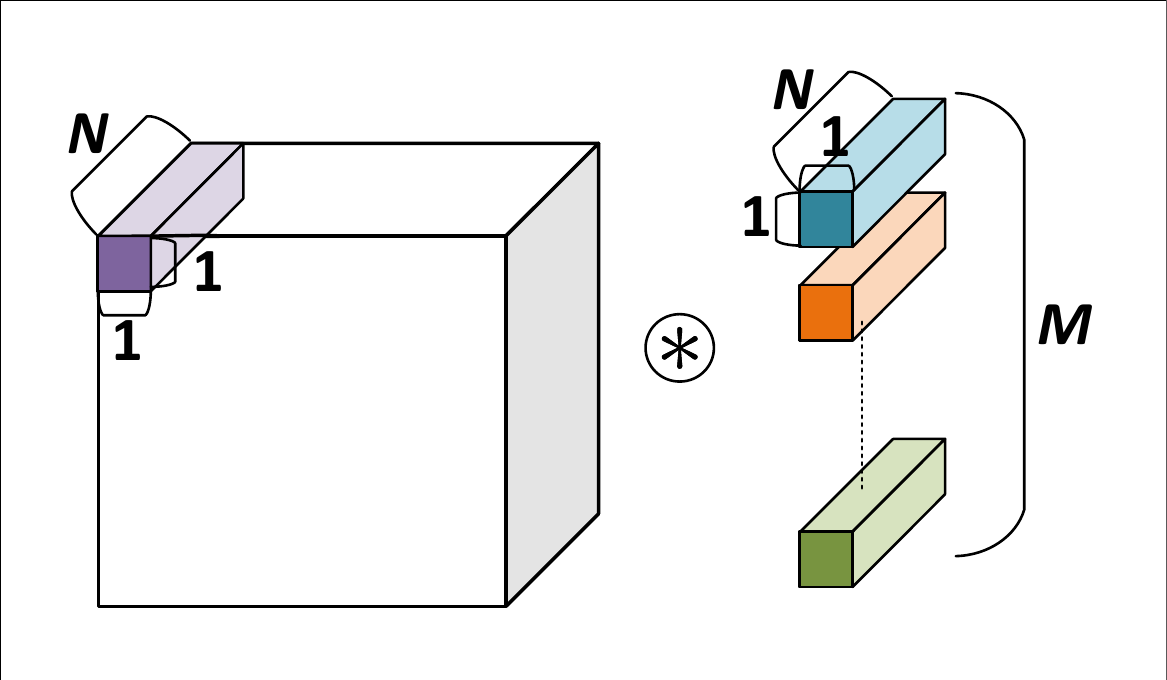}}%
	\label{fig:pw}%
	}
	\caption{Different convolutional operators.}
\end{figure}

Fig.~\ref{fig:cnn} shows normal \cn~filters with the shape of $M \times N \times K \times K$; thus, the computational cost of normal \cn~is $C = H\times W \times K^2 \times N \times M$. Group-\cn, shown in Fig.~\ref{fig:gc}, minimizes the computation cost of a \cn~operator by grouping its channel in $G$ receptions, reducing computation to $C/G$, where $M=f \cdot G~|~f \in \mathbb{N}$. Depthwise \cn \cite{mobilenet, xception} is an extreme case of group-\cn, where $G=N , f=1$. In this case, each filter is applied to each input channel individually based on Fig.~\ref{fig:dw}, and in contrast to the normal \cn, there is no reduction (summation) across channels. Pointwise \cn~is another type of operator which minimizes the computation by not capturing spatial dependencies within a frame pixels by setting the kernel size to $1\times1$.

\begin{figure}[tb]
    \centering
    \subfigure[Inverted Residual Block: MobileNet-V2]{
        \includegraphics[width=0.65\textwidth,trim= 18 15 18 20,clip]{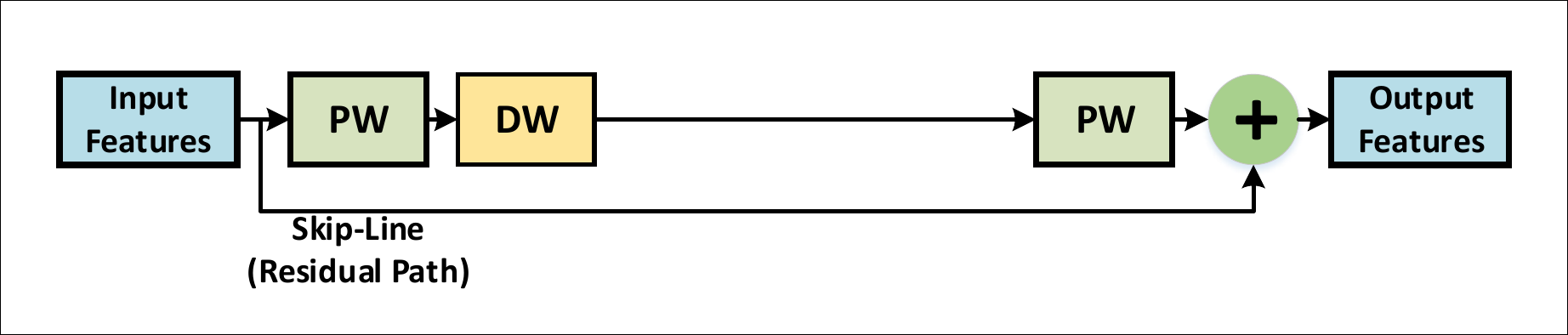}
    \label{fig:IRB_SKIP}
    }
    \\
    \subfigure[Inverted Residual Block: EfficientNet]{
        \includegraphics[width=0.65\textwidth,trim= 18 15 18 15,clip]{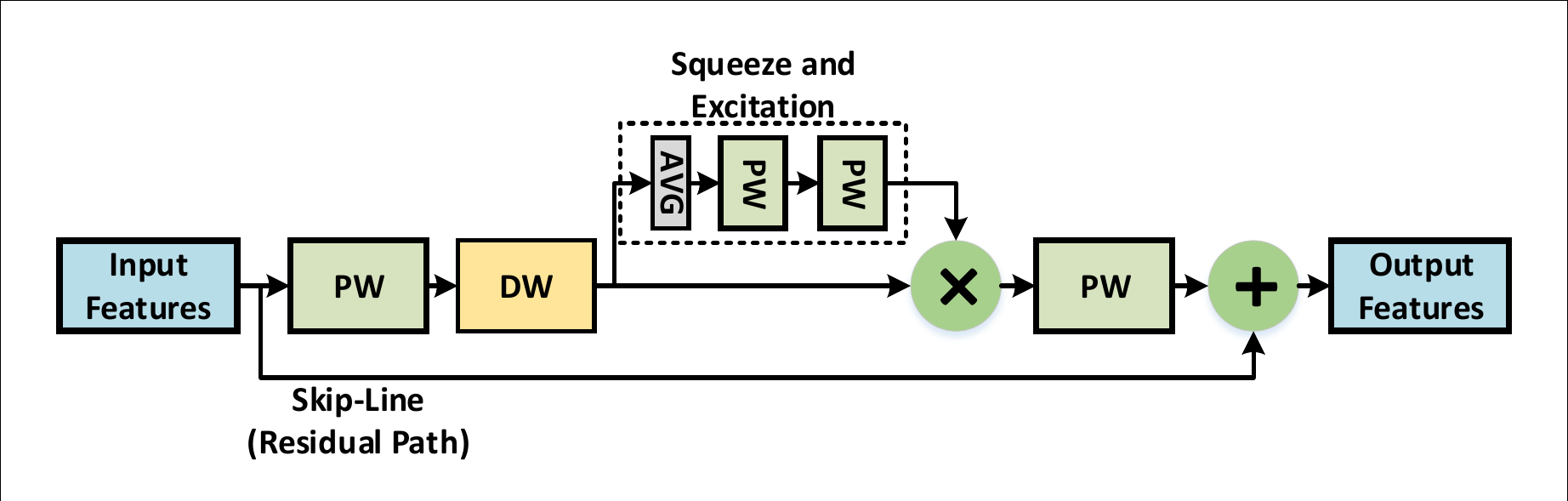}
    \label{fig:IRB_Effnet}
    }
    \caption{Inverted Residual Block (IRB) for \mn~(a) and \effnet~(b), respectively. The illustration of Batch Normalization and Activation layers repeated after each \cn~are ignored.}
    \label{fig:mn2_irb}
\end{figure}

As mentioned earlier, \dw~\cn~minimizes computation by removing reduction along the input channels; thus, it is not able to capture the channel-wise information. In the same fashion, \pw~\cn~reduces the computation complexity by removing spatial filtering, while it has a full reduction in channel depth. Depthwise separable \cn, used in \vmn~\cite{mobilenet}, is an integrated operator composed of a \dw~convolution, followed by \pw~\cn, in order to capture information in both spatial and channel domains, respectively. However, there is still information loss as features move along the network depth and are embedded into lower-dimensional space. \mn~\cite{mobileNetV2} introduced inverted residual connections to its previous network, further reducing both multiply-add operations, and model size, without sacrificing the network accuracy. The idea of residual connections was inspired by the ResNet \cite{ResNet} architecture to minimize information loss and speed up the training phase. Fig.~\ref{fig:mn2_irb} shows the structure of the Inverted Residual Block (IRB). IRB consists of a \pw~(expansion) \cn, followed by a \dw~\cn, followed by another \pw~(projection) \cn, to embed the features in a lower dimension. \red{The \mn~can control IRB layer input channel width, i.e., $N$, by altering the $\alpha$, which changes $N$ to $\alpha \times N$. The $\alpha=1$ is the baseline model. Selecting $\alpha < 1$ can reduces the computational complexity and the model size quadratically by roughly $\alpha^{2}$. We have examined the effect of this knobs and image input size on the final hardware performance and its accuracy in Section \ref{sec:ExpResults}.}

Another recently introduced example is \effnet, which further optimizes the IRB by adding Squeeze and Excitation (SE) blocks. Fig.~\ref{fig:IRB_Effnet} presents the \effnet~IRB with SE block. The SE block consists of a squeeze operation that captures the global spatial features, followed by an excitation operation that uses a gating function to allow important features to be captured while ignoring the rest. Traditionally, the normal sigmoid is used as the gating function for the SE block, but is replaced with the hard sigmoid to further reduce computation complexity. The hard sigmoid is a non-smooth approximation of the sigmoid function and is described as:
\begin{gather}
\frac{ReLU6(x+3)}{6},\\
ReLU6(x)= 
\begin{dcases}
    x,& \text{if } 0\leq x \leq 6\\
    0,              & \text{otherwise}
\end{dcases}
\end{gather}

\red{The design principles of DSCNNs result in relatively higher computational sparsity due to heterogeneous computing operators that cannot share hardware resources. Depthwise \cn~accumulates only across the spatial axis and needs only $K \times K$ fused-multiply-add (FMA) operations since its weight shape is $[M, 1, K, K]$. Since versatile systolic arrays are often designed to support both spatial and channel accumulation, they perform more FMA operations. They map \dw~to matrix multiplication problem by kernel zero-padding and reshaping appropriately; however, the cost of memory real estate, and the redundant computation demand, are not affordable for resource-constrained hardware platforms.} 

In next, we introduce DeepDive as a fully vertical and versatile solution to support sparse operators introduced in DSCNNs. As case studies, we selected \mn~and \effnet~as two examples of DSCNNs, and we thoroughly elaborate their implementation with the aid of DeepDive in section \ref{subSec:CaseStudy} and \ref{subSec:CaseStudy_effnet}, respectively.

\begin{figure}[t]
\centering
\includegraphics[width=.5\textwidth, trim= 25 15 15 20,clip, keepaspectratio]{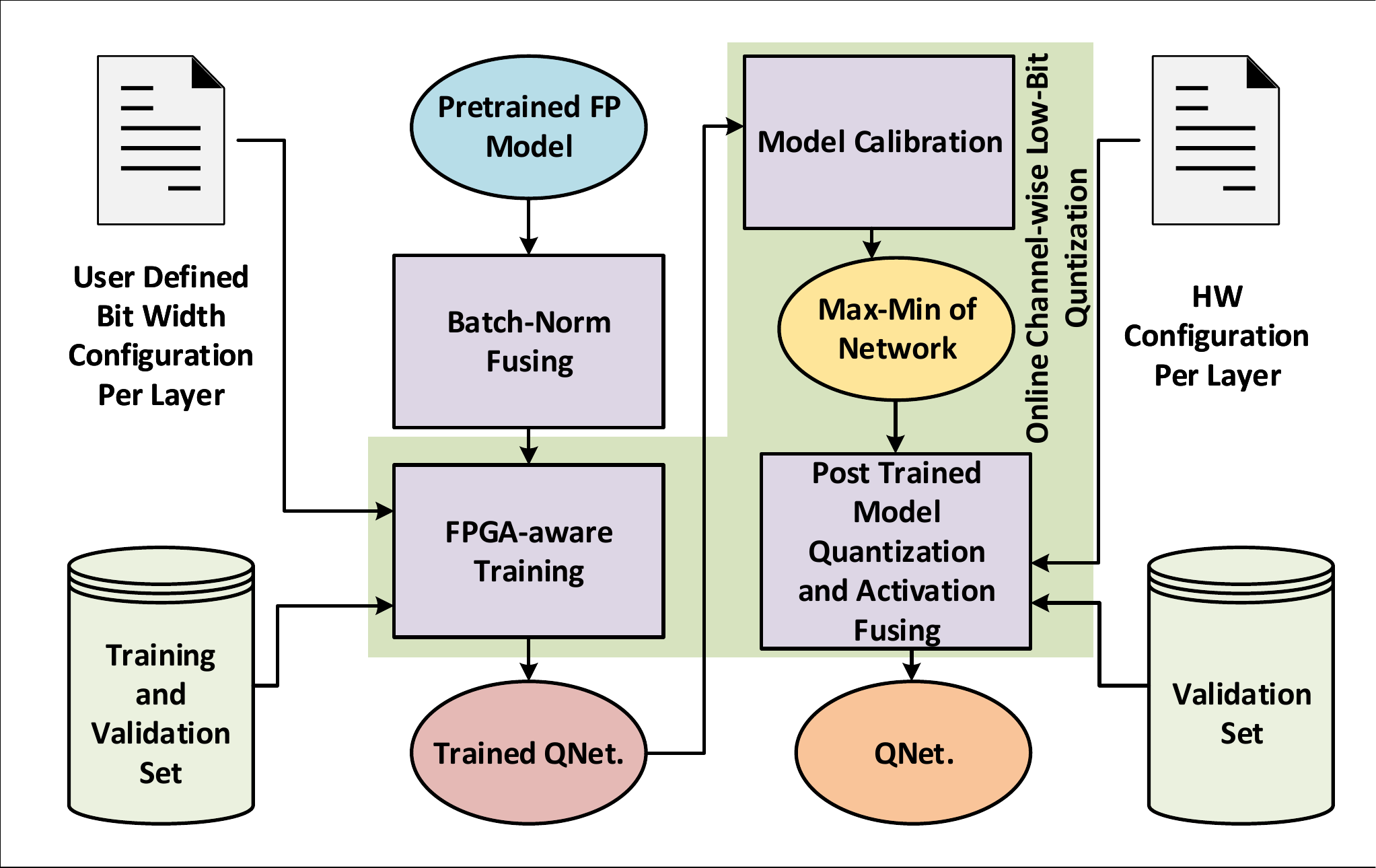}
\caption{DeepDive: Front-end.}
\label{fig:DeepDive_FrontEnd}%
\end{figure}

\section{DeepDive: Front-end}\label{sec:DD_F}
This section describes the front-end of DeepDive, which brings hardware-awareness into training DSCNNs.  Fig.~\ref{fig:DeepDive_FrontEnd} illustrates the main components of the front-end and their corresponding output. The procedure starts by feeding a pre-trained floating-point network into the DeepDive. The \textit{Batch-Norm Fusing} merges the batch-normalization operator into the \cn~in order to remove any floating-point operations in the final hardware solution. Next, \textit{Online Channel-wise Low-Bit Quantization} quantizes while training the fused network at extremely low-bit resolutions (e.g., 3-6 bit) across all channels within separable layers. Then, the trained network will be calibrated by extracting the minimum and maximum values across all channels per layer of the network. The \textit{Post-Trained Model Quantization} then uses these acquired ranges to fuse the activation layer, i.e., ReLU6, into the \cn~operator. The outcome, \textit{QNet}, consists of only \cn~operators that have had their output set to the minimum and maximum quantized value automatically---when they are less than 0 and greater than 6, respectively. In the following, we explain the details of two important aspects of front-end: (1) Batch-Norm Fusing, and (2) Online Channel-wise Low-Bit Quantization.

\subsection{Batch-Normalization Fusing}
Batch-Normalization (BN) \cite{BatchNorm} is a linear operator, generally seen following a \cn~layer, in order to normalize the output of the \cn. BN improves the training speed and stability of the network. The BN function is defined by Eq.~\ref{eq:bn}:
\begin{equation}
    \reallywidehat{x} = \gamma \dfrac{x_j - \mu}{\sqrt{\sigma^2+\epsilon}} + \xi,
    \label{eq:bn}
\end{equation}
where $\gamma$ is BN weight, $\xi$ is its bias, and $\mu$, and $\sigma$ are mean and variance of training batch calculated during the training, respectively. $\epsilon$ is a small constant defined to prevent division by zero. Both $\gamma$ and $\beta$ are trainable parameters. For networks where its \cn~operators are always followed by BN, DeepDive online training fuses these two consecutive layers together by applying following equations:
\begin{gather}
    \reallywidehat{v} = {(\sigma^2+\epsilon)}^{\frac{-1}{2}},\\
    \reallywidehat{\omega}_{conv} = \omega_{conv} \times diag(\gamma \cdot \reallywidehat{v}),\\
    \reallywidehat{B}_{conv} = B_{conv} + (\xi - (\gamma\cdot\mu\cdot\reallywidehat{v})),
\end{gather}
where $\omega_{conv}$ and $B_{conv}$ are trained weights and biases of \cn~operator, respectively. After BN fusion, the network model is ready for quantize-aware training.

\subsection{Online Channel-wise Low-bit Quantization}
Quantization is a well-known approach to compress the network model size, and speed up the computation, by mapping number representations from floating-point single precision (FP32) to integer representation. Due to the malleability of FPGA fabrics, designers can greatly reduce the integer bit-width, while minimizing the introduced quantization error, by training the network for the new representation. DeepDive applies the Range-Based Linear quantization to compress the network weights and biases. Let's define $\mathbb{T}=\{x~|~x \in \mathbb{R}\}$, such that $\mathbb{T}$ is the floating-point pre-trained network model. Function $h:~ \mathbb{T}~\rightarrow~\mathbb{Q}$ will map and scale $\mathbb{T}$ to $\mathbb{Q}$, where $\mathbb{Q}$ is quantized integer representation set. Eq.~\ref{eq:LQ} defines function $h$:
\begin{equation}\label{eq:LQ}
    x = S(x_{q}+m_{zp})~|~x_{q}, m_{zp} \in \mathbb{Q},
\end{equation}
where $S\in\mathbb{R}$, is the scaling factor, $x_q$ is the quantized value, and $m_{zp}$ is the zero-point defined to make the right-hand side of Eq.~\ref{eq:LQ} equal zero when $x_{fp}=0$. Based on the range of $x_q$, two methods of Asymmetric Representation and Symmetric Representation are defined. In asymmetric mode the $min_{x}=min(x)$ is mapped to 0, while $max_{x}=max(x)$ is $2^{BW}-1$, while $BW$ is the bit-width.  In contrast, symmetric maps both $[min_x$, $max_x]$ to $[-(2^{BW-1}),~2^{BW-1}-1]$. \mn~uses ReLU6 as its non-linearity function --- its output is always positive and less than 6. Therefore, we opted for the asymmetric method, since the negative range of the symmetric representation is not useful, and we are not able to benefit from the full range of representation; thus, it will have an impact on the output accuracy of each activation layer.

\begin{figure}[tb]
	\centering
	\includegraphics[width=.5\textwidth, trim= 20 15 20 18,clip, keepaspectratio]{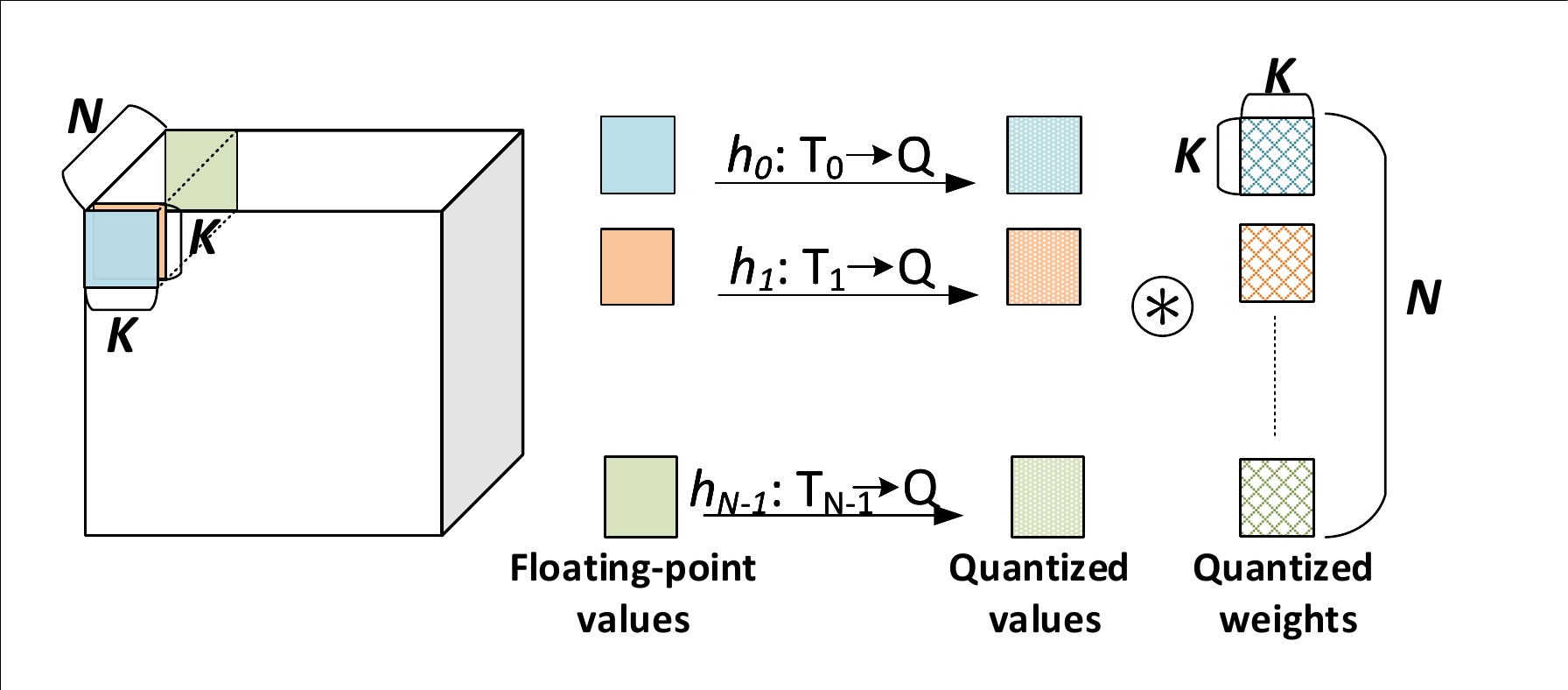}
	\caption{Per-channel range-based linear quantization. In this \dw~\cn~example, per each $N$ output channel, a separate mapping function is created.}
	\label{fig:linearRange_asym}
\end{figure}

DeepDive can quantize a network model per output channel, or per \cn~layer. Per layer approach defines $h$ function per whole \cn~layer, while per-channel quantization defines $h_{j}~|~j=0, \cdots, M-1$ per each output channel for a \cn~operator. For instance, Fig.~\ref{fig:linearRange_asym} shows the per-channel quantization approach for a \dw~\cn. 

After the network is trained and quantized based on the user-provided configuration, the validation set is used again for the network model calibration. The calibration data will be used to make the trained network ready for post-training quantization. In this step, based on the acquired min-max, and the type of quantization, the scaling $S$ and $m_{zp}$ will be recalculated again to re-evaluate $h_{j}$, which results in $h^{pq}_{j}: [0,~6] \rightarrow [0,~2^{BW}-1]$. By applying this approach, DeepDive fuses the ReLU6 activation to the \cn~operator. 

\begin{figure}[t]
\center
\includegraphics[width=.5\textwidth, trim= 18 17 17 17,clip, keepaspectratio]{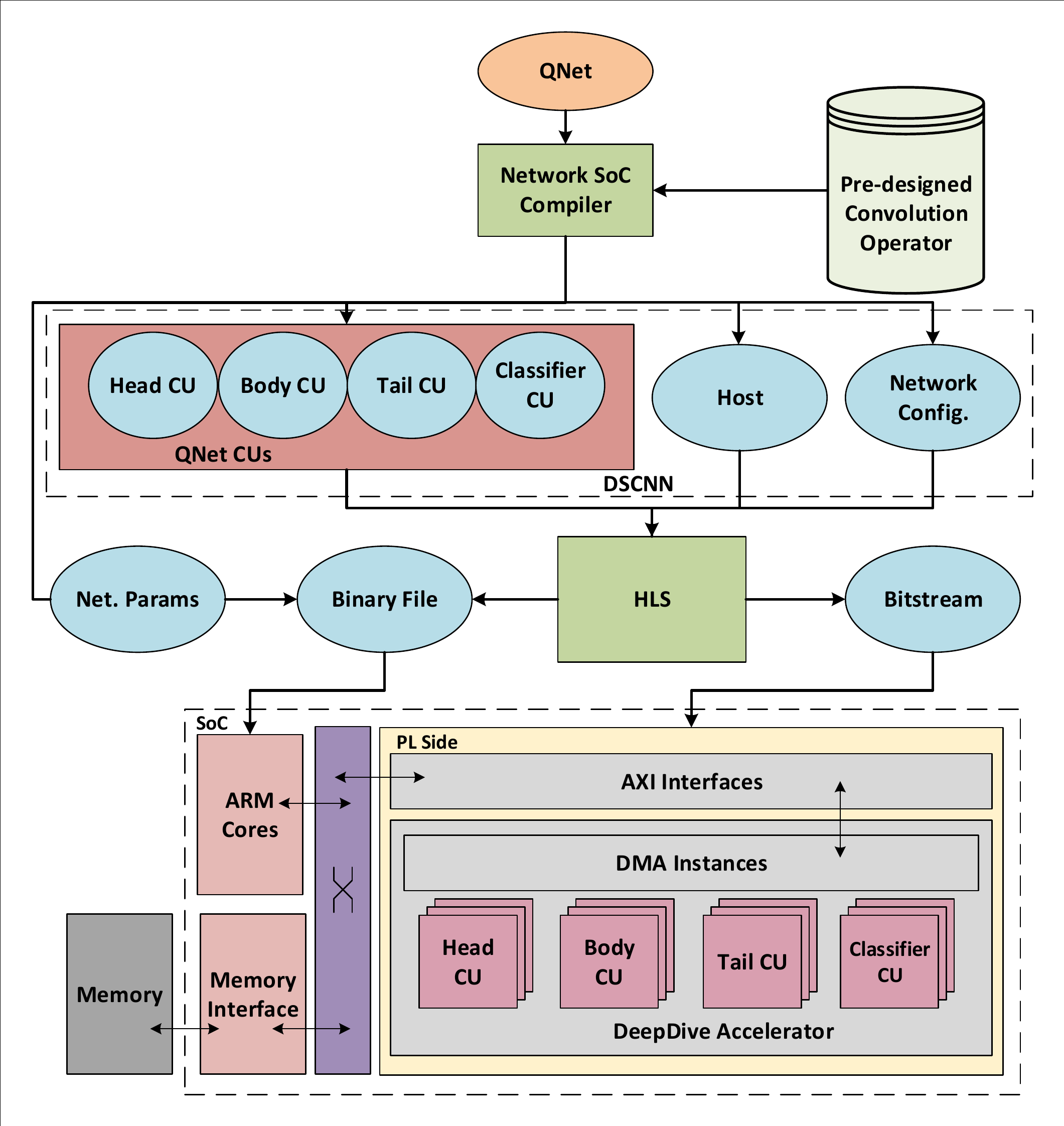}
\caption{DeepDive: Back-end.}
\label{fig:Backend}%
\end{figure}

\section{DeepDive: Back-end}\label{sec:DD_B}
DeepDive's back-end offers a novel micro-architectural approach, and design flow, customized for efficient execution of DSCNNs on edge FPGAs. Fig.~\ref{fig:Backend} presents the DeepDive back-end design flow. The heart of DeepDive's back-end is the \textit{Network SoC Compiler}. It receives the design properties from DeepDive's front-end and generates a full design of the system for both hardware (as synthesizable C++ models mapped to FPGAs fabric), software codes, and system configurations. To generate the optimized hardware for DSCNNs, the Network SoC Compiler \red{uses pre-designed highly-optimized RTL micro-architectural blocks or synthesizable C++ model for \dw, \pw, and normal \cn~operators}. In simple words, the Network SoC Compiler generates a network graph containing the network layout and data dependencies. It then creates key heterogeneous CUs, called \textit{QNet Accelerators}, with respect to DeepDive's system architecture. 

In the following, at first, we describe micro-architectural details of convolutional operators, and then we discuss the details of the Network SoC compiler and system architecture.



\begin{figure}[tb]
\center
\includegraphics[width=\linewidth, trim= 10 10 10 10,clip, keepaspectratio]{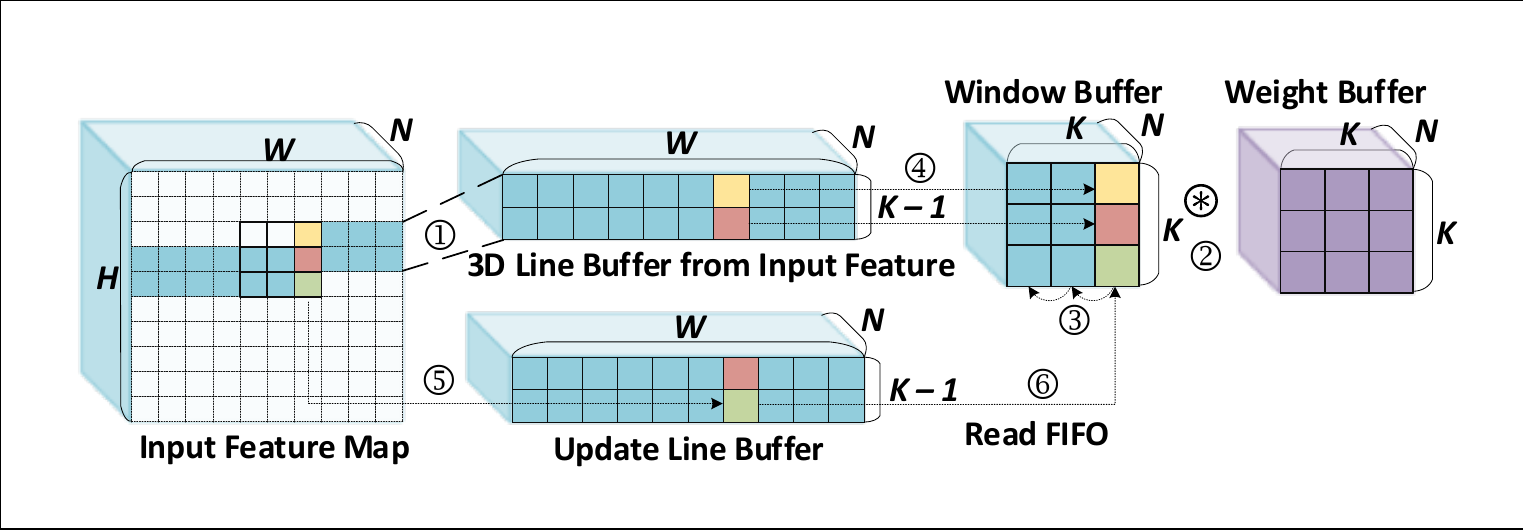}
\caption{\red{Shift and update\cmmnt{The data movement and update} mechanism of Window and Line Buffer.  \textcircled{\raisebox{-0.9pt}{1}} Line Buffer is filled with input feature data. \textcircled{\raisebox{-0.9pt}{2}} Window Buffer is convoluted with weights. \textcircled{\raisebox{-0.9pt}{3}} The data in window is left shifted.   \textcircled{\raisebox{-0.9pt}{4}} New data from the line buffer is copied in to the window. \textcircled{\raisebox{-0.9pt}{5}} \& \textcircled{\raisebox{-0.9pt}{6}} Data from the FIFO is then copied into the line buffer and window buffer. All the Data Movements are pipelined.}}
\label{fig:LineBuff}%
\end{figure}

\subsection{Convolutional Operators}\label{subSec:Conv_op}
Since DeepDive is specially designed for DSCNNs, it naturally supports all convolutional operations, namely, normal \cn, \dw~\cn, and \pw~\cn. Each convolution operator buffers minimum job data size, which is necessary to start the computation, with the assumption that the network parameters necessary for computing are transferred to internal memory, and that the intermediate feature maps are streamed in and out. These operators are pipelined and parallelized in a way that is ideal for both memory-bound and compute-bound operations. The heart of a convolutional operator is a reconfigurable \textit{Direct Convolution} core with different degrees of parallelism. The amount of parallelism defines the utilization, and parallel read/write ports required by the scratchpad or local buffers. This flexibility allows the Network SoC Compiler to manage the resources efficiently by tweaking the parallelism knobs to achieve the best performance (will be further discussed in section \ref{sec:soccompiler}). Next, we elaborate on each operator from the design standpoint. In addition, we formulate the amount of parallelism per each convolutional operator.

\begin{figure}[h]
\center
\includegraphics[width=.5\textwidth, trim= 25 30 18 35,clip, keepaspectratio]{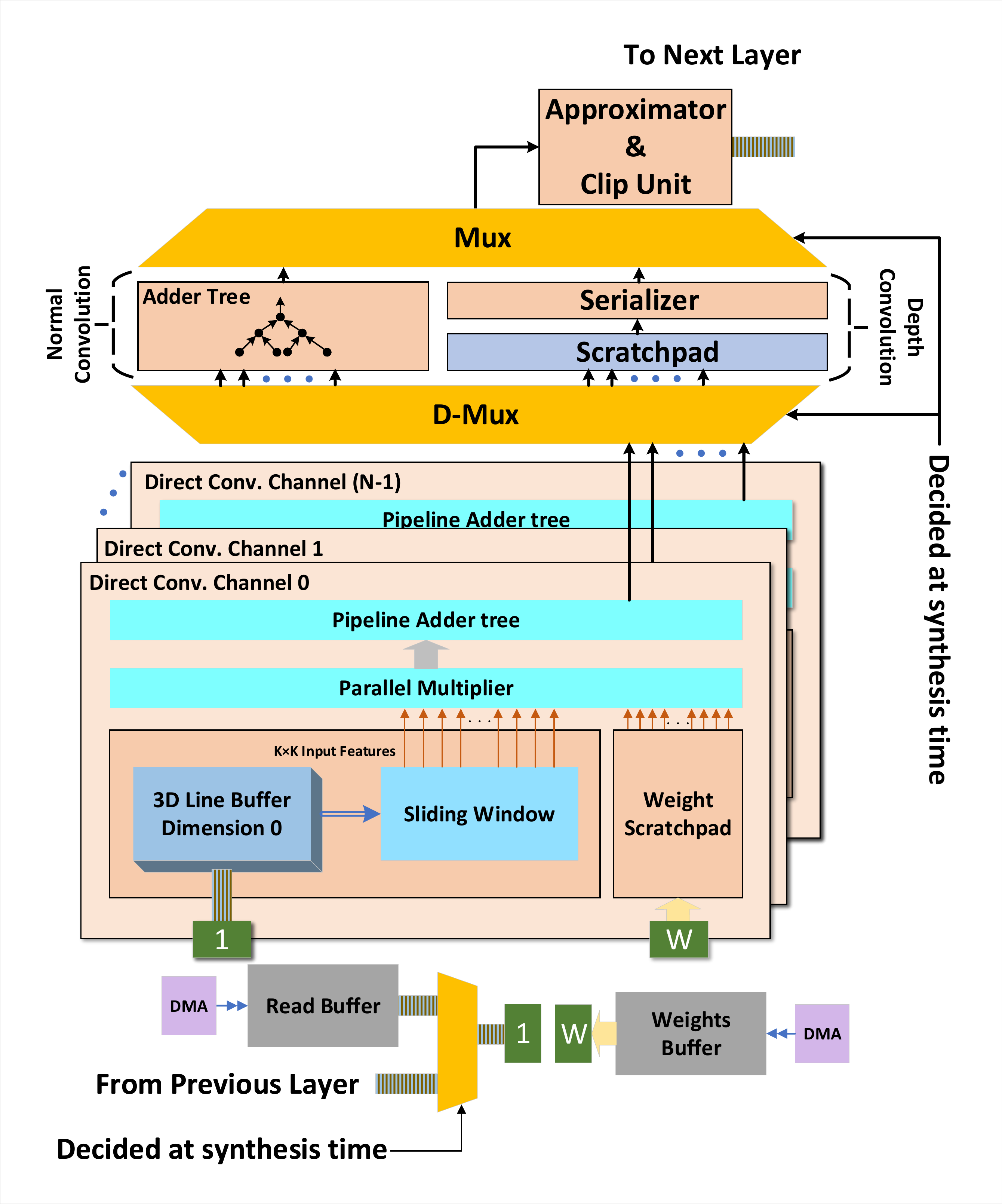}
\caption{Schematic block diagram of \dw~\red{and normal \cn.}}
\label{fig:DW_RTL}%
\end{figure}

\subsubsection{Depthwise Convolution}
The Depthwise convolution uses a 3D line buffer and 3D window to perform direct convolution. The input feature is streamed into a line buffer and then copied into a window buffer with parallel read access, as shown in Fig.~\ref{fig:LineBuff}. Once the computation is finished, the data in the computation core will be flushed and reloaded with the new one from the line buffer. The hardware design ensures the data movement involved in this process is fully pipelined, and the initiation interval is limited to a single cycle. Computation starts as soon as the required amount of data is streamed from the main memory. For the current design, the max achievable parallelism is limited to the $K$ and $N$. 

Fig.~\ref{fig:DW_RTL} presents the micro-architecture of \dw~and normal \cn~operators. As depicted in Fig.~\ref{fig:DW_RTL}, the selected input is read in streaming fashion into the 3D line buffer and then copied into the sliding window. The weights are burst read into the weight scratch pad. The Sliding Window and the Weight scratchpad have multiple read ports. Every channel of the input is processed by the direct convolution compute core. The direct convolution compute core has a parallel multiplier, and a pipelined adder tree, together which carryout the MAC operation, followed by the Approximator and Clip unit. This unit truncates, or rounds, the results and then clips them to $[0,~2^{BW}-1]$ based on the quantization parameters extracted at the front-end for this operator. Therefore, this unit also acts as the ReLU6 activation layer defined in \mn~or \effnet. The \dw~\cn~is more sparse, and has the least amount of data reuse. The maximum parallel operations are calculated as the following:
\begin{gather}
\label{eq:dw_par}
    Parallel Ops = K^{dw}_{max} \times K^{dw}_{max} \times N^{dw}_{max},
\end{gather}
In Eq.~\ref{eq:dw_par}, $K^{dw}_{max}$, and $N^{dw}_{max}$ are the maximum kernel size and maximum input-channel across all the \dw~convolutions in the network, respectively. 

\subsubsection{Normal Convolution}
\red{The DSCNN has one normal \cn, and it is the first operator to embed patterns from both spatial and channel dimensions from the given input image. Since the next layer after normal \cn~ is \dw, it is essential to generate output pixels column-wise (spatial dimension) so the \dw~can start the job immediately. Therefore, we improve the parallelism level by having a dedicated adder tree located after the direct convolution kernel for the input channel reduction. The block diagram of normal \cn~is, also shown in Fig.~\ref{fig:DW_RTL}.} The parallelism in normal \cn~is across kernel size and input channels --- described in the following:
\begin{gather}
\label{eq:cn_par}
    Parallel Ops = K^{nc}_{max} \times K^{nc}_{max} \times N^{nc}_{max},
\end{gather}
where $N^{nc}_{Max Size}$ is the maximum input channel size, and $K^{nc}_{max}$ is the maximum kernel size, assigned from all normal \cn. Normal \cn~has slightly more data movements compared to the \dw~\cn~due to the pipelined adder tree implemented at the end of direct \cn~core.


\begin{figure}[tbh]
\center
\includegraphics[width=.5\textwidth, trim= 20 20 20 25,clip, keepaspectratio]{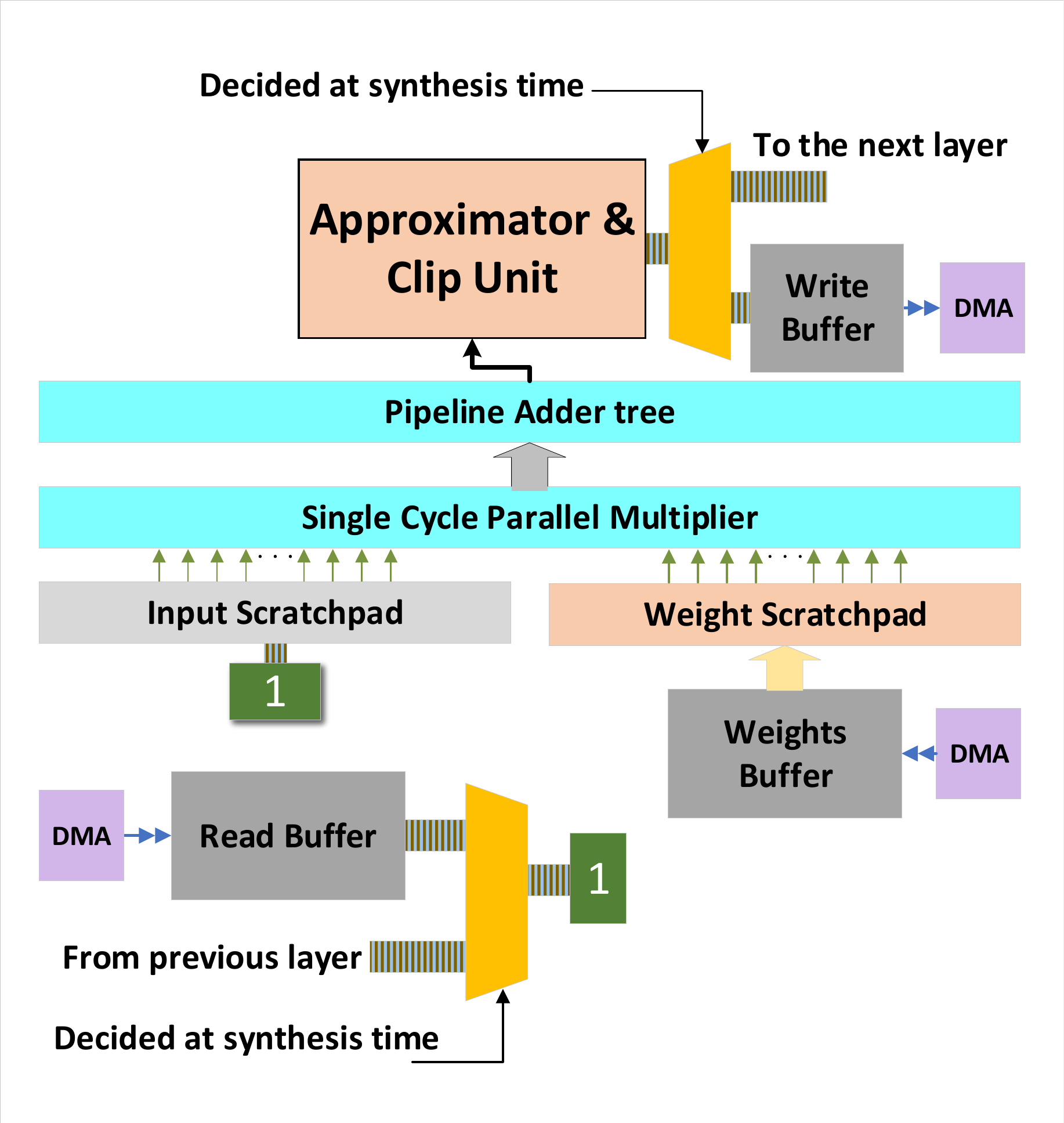}
\caption{Schematic block diagram of \pw~\cn.}
\label{fig:pw_RTL}%
\end{figure}

\subsubsection{Pointwise Convolution}
Due to the dense operation of \pw, the design of this operator can be similar to the design of a general matrix multiplication, which is well suited for the systolic array. With maximum data reuse, this operator can leverage maximum parallelism. It has both fewer algorithmic, and fewer data movement complexity, which makes it best fit for a high amount of parallelism. Fig.~\ref{fig:pw_RTL} shows the structure of \pw~convolution operator. The required input is directly read into the input scratchpad from the read buffer. The weights are burst read into the weight scratchpad. The input buffer and the weight scratchpad have multiple read ports for parallel data access. The single-cycle parallel multiplier and the adder tree take advantage of the multiple ports to perform the MAC operations in  parallel fashion. The amount of parallelism for our design is across the input channels.
\begin{gather}
\label{eq:pw_par}
    Parallel Ops = N^{PW_{type}}_{max},
\end{gather}
where $N^{PW_{type}}_{max}$ is the maximum input channel size across all the specific $type$ (eg. projection or expansion \pw~in the \mn) of \pw~convolutions mapped to specific compute unit.



\subsection{Network SoC Compiler}
\label{sec:soccompiler}
The Network SoC Compiler observes the network graph, the targeted hardware device, and existing pre-designed synthesizable C++ IPs for \cn, and then translates the network graph by grouping the convolutional operators into customized \qnet~CUs with respect to system architecture. It tweaks the hardware architectural knobs to maximize parallelism, fusing as many convolutional operators as possible to reduce the number of shared memory transactions, and increase the overlap between computation and memory latency. Based on the repetitive pattern, it wraps the \cn~operators in four different \htr~CUs: \textcircled{\raisebox{-0.9pt}{1}} The \textit{Head CU} generally consists of \nc~followed by a special case of IRB which is only called once; \textcircled{\raisebox{-0.9pt}{2}} The \textit{Body CU} invokes IRB since it has maximum repetitions based on the DSCNNs architectures; \textcircled{\raisebox{-0.9pt}{3}} The \textit{\raisebox{-0.9pt}Tail CU} usually consists of \pw~\cn~followed by Average Pooling to embed the features and make them ready in respect of size and shape for the classifier; \textcircled{\raisebox{-0.9pt}{4}} Finally, the mapping of Tail CU output to $k-$classes is accomplished by \textit{Classifier CU}. 

Below, we describe the details of Network SoC Synthesizer including, system architecture, memory organization, Heterogeneous \qnet~CUs, host code scheduling and CUs management.



\begin{figure}[h]
\centering
\includegraphics[width=.5\textwidth, trim= 20 20 15 20,clip, keepaspectratio]{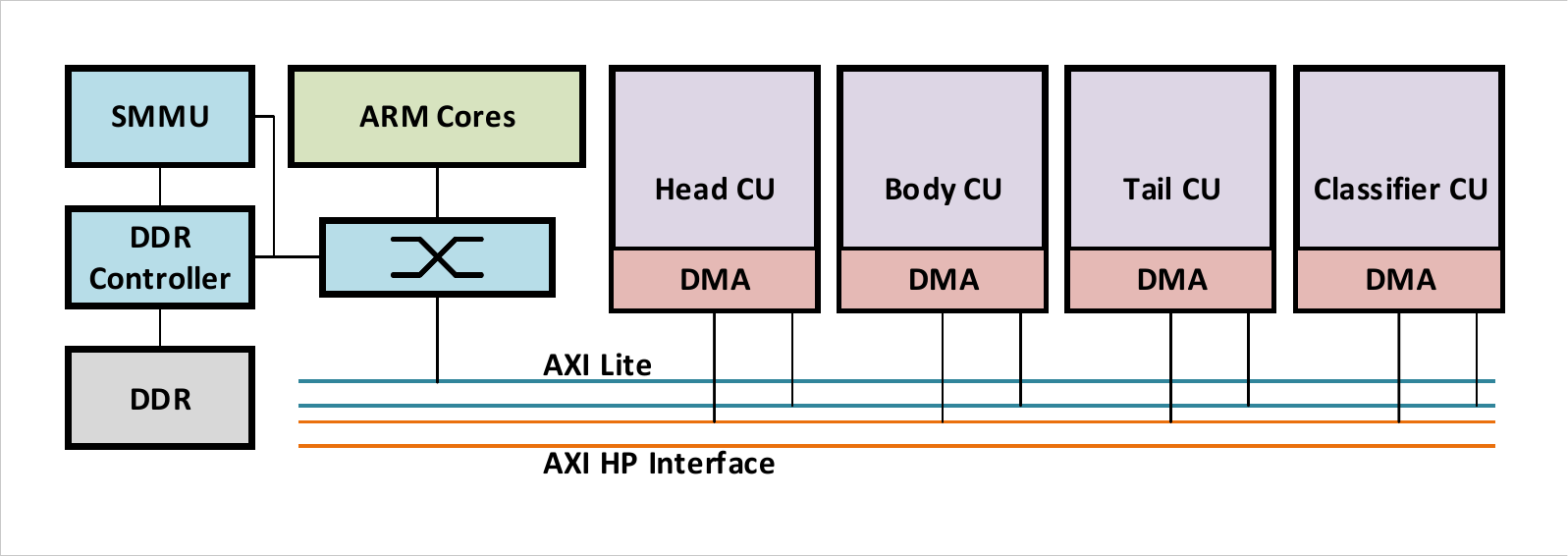}
\caption{System level architecture of DeepDive.}
\label{fig:system_arch}%
\end{figure}

\begin{figure*}[t]
	\centering
	
	\subfigure[Head Computing Unit]{%
		{\includegraphics[width=0.45\linewidth, trim= 20 20 20 20,clip, keepaspectratio]{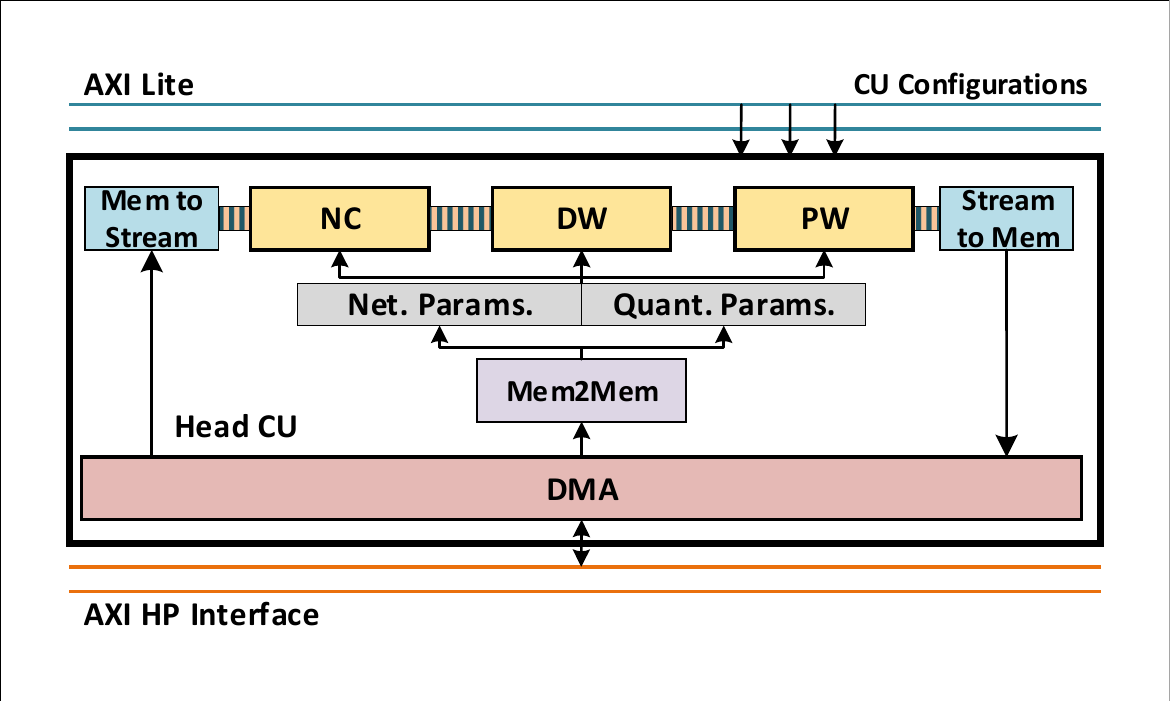}}%
		\label{fig:Net_Head}%
	}\qquad
	\subfigure[Body Computing Unit]{%
		{\includegraphics[width=0.45\linewidth, trim= 20 20 20 20,clip, keepaspectratio]{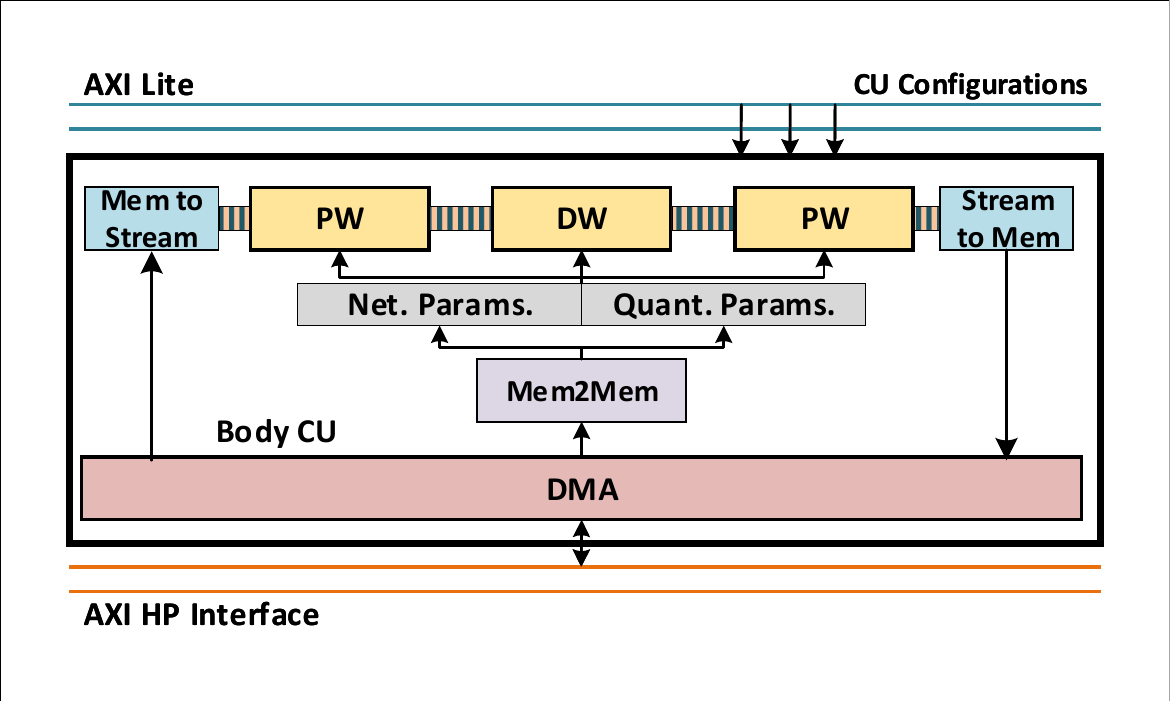}}%
		\label{fig:Net_Body}%
	}\\
	\subfigure[Tail Computing Unit]{%
		{\includegraphics[width=0.45\linewidth, trim= 20 20 20 20,clip, keepaspectratio]{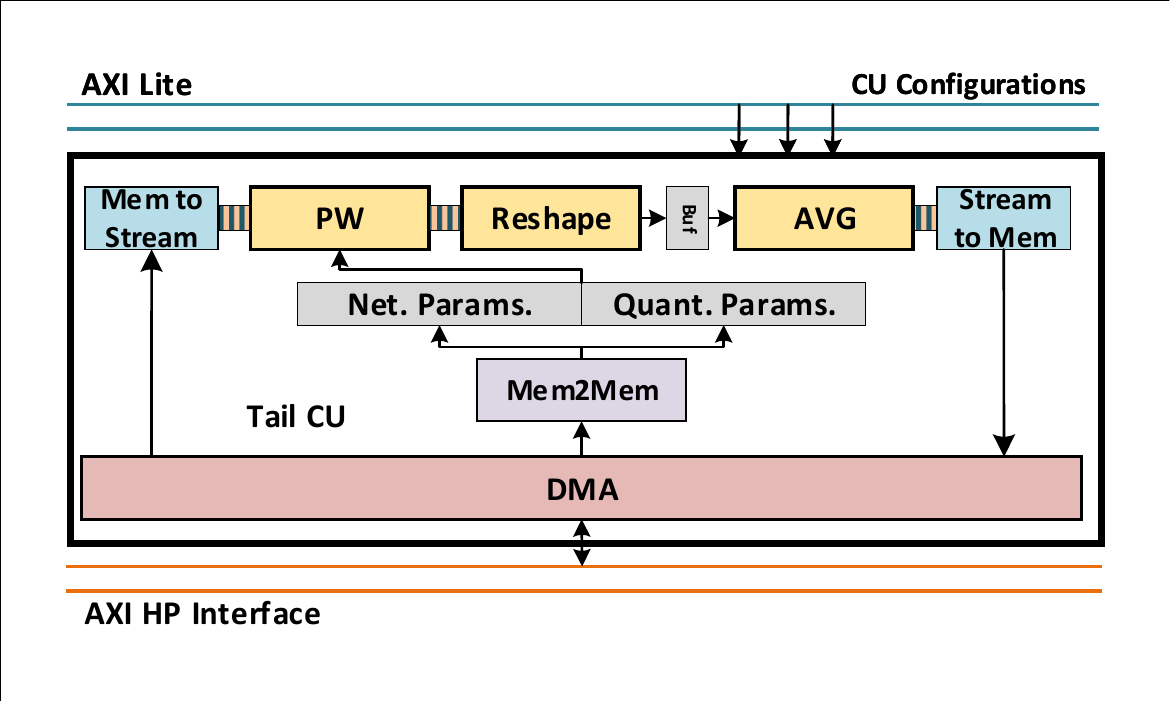}}%
		\label{fig:Net_Tail}%
	}\qquad
	\subfigure[Classifier Computing Unit]{%
	{\includegraphics[width=0.45\linewidth, trim= 20 20 20 20,clip, keepaspectratio]{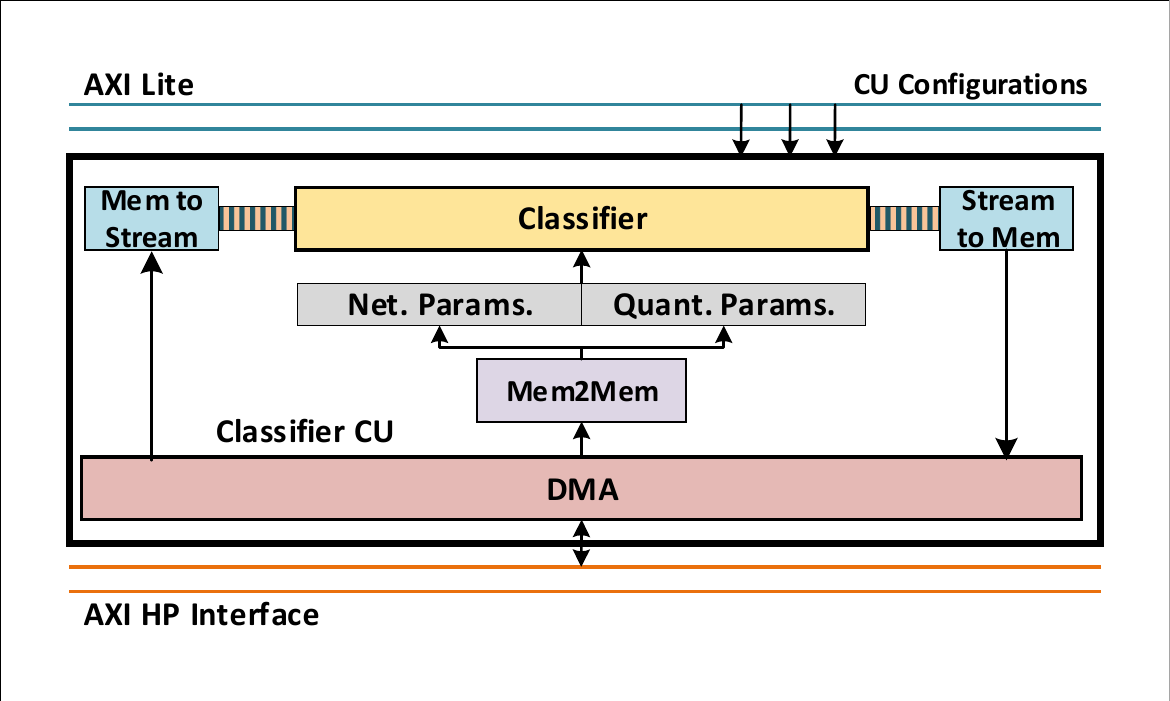}}%
	\label{fig:Net_Classifier}%
	}
	\caption{Architecture of \qnet~Heterogeneous Computing Units for \mn.}
\end{figure*}

\subsubsection{DeepDive System Architecture}
\label{subsec:Hetro_Compu_unit}
As emphasized before, the convolutional operators of DSCNNs demonstrate a repetitive structural behavior wherein some either appear once, or they are repeated across the entire network. Depending on the recurrence of the convolutional operators, they are mapped to the Head, Body, Tail, and Classifier CU. Fig.~\ref{fig:system_arch} shows the system architecture of DeepDive Hardware Accelerator. Each CU has its own dedicated Direct Memory Access (DMA), and its parameters, such as array pointers, $N$, $M$, and $H$, can be configured at runtime via the control bus (e.g., AXI Lite Bus). After configuration, each CU can transfer the input/output features map and weights tensors via streaming channels (e.g., AXI HP Interface) \red{through System Memory Management Unit (SMMU)}. The composition of CU is parameterized by the buffer shapes, data type widths, and the computation core, which are a few of the architectural knobs provided while designing the hardware accelerator. This makes our design scalable and reconfigurable for DSCNNs. We will discuss our hardware knobs and each CU's internal composition in detail after we explain the memory transactions and management. The CUs are scheduled and pipelined to increase the concurrency. 



\subsubsection{Memory Organization} 
\label{sssec:MemM_T}
Each CU has its own dedicated buffer and scratchpad to handle its memory requirements. The memory layout of the on-chip buffers are designed to satisfy the data access pattern required by the convolutional operators, in order to minimize the pipeline depth implemented in the computation core. The memory transactions in the CUs can be categorized into two groups: \textcircled{\raisebox{-0.9pt}{1}} memory to memory transaction, where data is burst read from DDR memory to PL memory, and \textcircled{\raisebox{-0.9pt}{2}} memory to stream transaction, where data is streamed via DMA to or from PL memory. As an example, Fig.~\ref{fig:Net_Head} demonstrates the memory transactions for Head CU targeted for \mn. Convolutional network parameters like weights, quantization parameters, and biases are burst read from DDR to PL buffers. The input/output feature maps are streamed from DDR to PL. Apart from memory transactions of input/output features between DDR and PL, the inter-CU data transfers within its operators also occurs in streaming fashion, where intermediate feature map data is streamed in-between different convolutional layers. Stream FIFO offers two main advantages, memory and computation latency overlap and data movement reduction between DDR and PL.

\subsubsection{QNet Heterogeneous CUs}
In this subsection, we will explain the \htr~CUs, and the available architecture knobs that can be tweaked based on hardware and performance constraints. As mentioned earlier, Network SoC Compiler creates four unique CUs for each DSCNNs. The CUs are completely parameterizable, and customizable, for scalability and flexibility. Following section describes each CU in detail. We also provide illustrative figures for the example of \mn.

\textbf{Head CU:} DSCNNs tend to start with a particular pattern, which comprises of a fixed set of layers that are not recurrent in any other part of the network. As explained in the section~\ref{sssec:MemM_T}, the Head CU has its own dedicated internal memory for buffers. The data transactions occur in memory-to-memory mode and the intermediate data streams between convolutional layers within the head CU. As an example, Fig.~\ref{fig:Net_Head} demonstrates the Head CU for \mn~model, which is composed of normal \cn~followed by \dw~and \pw~\cn, all fused by FIFO stream. This CU is scheduled once during the course of any DSCNN implementation. After running the head of CU, the repeatable pattern will be merged and mapped to the Body CU explained in the next part.

\textbf{Body CU:} The Body CU is the most important CU within DeepDive's system architecture. It is responsible for executing majority of DSCNNs blocks iteratively. As an example, the IRB, which is the most repetitive block of \mn, is entirely mapped to the Body CU. The IRB consists of \pw~(expansion), \dw, and \pw~(projection) layers, all running concurrently in a fused fashion within the Body CU. Fig.~\ref{fig:Net_Body} shows the structure of this CU for \mn. Upon examining the network graph of DSCNNs, we see that occasionally, the IRB needs to perform residual connections. Depending upon the network graph, DeepDive facilitates residual connections implementation within or outside the PL targeted device resources. The Body CU is parameterized so as to support both memory-bound IRBs, which ideally are earlier blocks of DSCNNs, and compute-bound IRBs, which tend to be later blocks of DSCNNs. Therefore, the network SoC compiler configures the Body CU with maximum buffer size needed by memory-bound IRBs, and maximum level of parallelism to meet the demand imposed by compute-bound IRBs. At the same time, the Body CU supports \cn~operations with variable stride over different IRBs. These features increase the framework inclusiveness by supporting multiple IRB scenarios within the same DSCNN.

\textbf{Tail CU:} The Tail CU consists of the last layers of DSCNNs. The task of this CU is to make the embedded feature size ready for the dense layer implemented in the Classifier CU. Fig.~\ref{fig:Net_Tail} represents the structure of Tail CU in \mn. This CU is comprised of a single \pw~\cn~operator, followed by an average pool. As intermediate feature maps are streamed from layer to layer in a channel-wise fashion, the reshape block reorders the memory layout of the feature map in a column-wise mode. Therefore, the average pooling can accumulate the input on-the-fly and stream out.

\begin{figure}[h]
\centering
\includegraphics[width=.5\textwidth, trim= 30 15 15 20,clip, keepaspectratio]{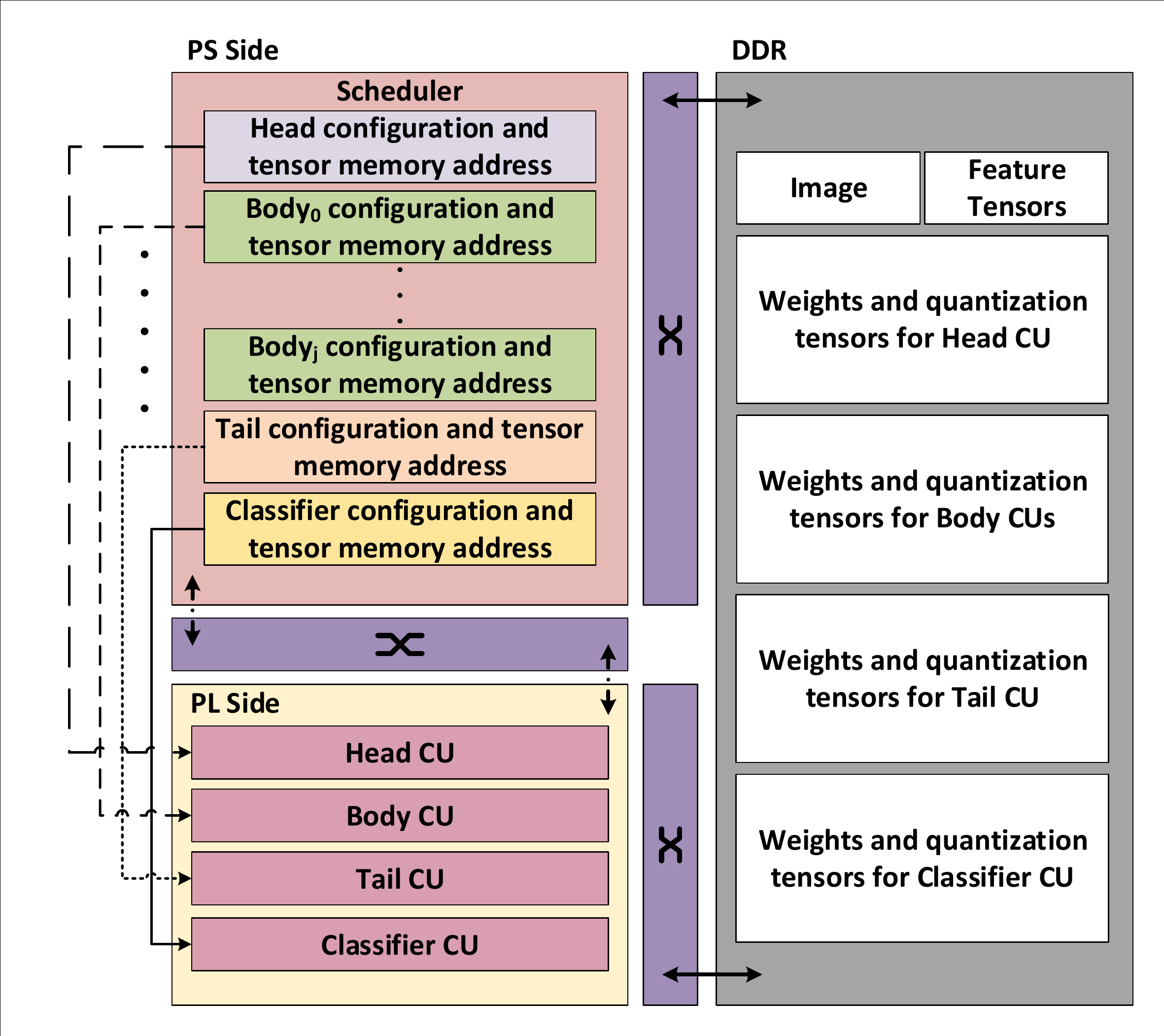}
\caption{Host level scheduling and memory footprint of CUs.}
\label{fig:runtime}%
\end{figure}

\begin{table*}[t]
\centering
\caption{Effect of altering $\alpha$ and $H$ for fixed $BW=4$}
\begin{adjustbox}{width=\linewidth,center}
  \begin{tabular}{lccccc|ccccc|ccccc|ccccc}
    \toprule
    \toprule
    $\alpha$ & \multicolumn{5}{c}{1}                 & \multicolumn{5}{c}{0.75}              & \multicolumn{5}{c}{0.5}               & \multicolumn{5}{c}{0.35} \\
    \midrule
    $H$     & 224   & 192   & 160   & 128   & 96    & 224   & 192   & 160   & 128   & 96    & 224   & 192   & 160   & 128   & 96    & 224   & 192   & 160   & 128   & 96 \\
    Params(Mb) & 13.31 & 13.31 & 13.31 & 13.31 & 13.31 & 10.01 & 10.01 & 10.01 & 10.01 & 10.01 & 7.48  & 7.48  & 7.48  & 7.48  & 7.48  & 6.37  & 6.37  & 6.37  & 6.37  & 6.37 \\
    \#Ops(M) &  313.621  &  230.755  &  160.638  &  103.269  &    58.649  &  220.326  &  162.212  &  113.038  &    72.805  &    41.513  &  104.164  &    76.868  &    53.772  &    34.875  &    20.177  &    64.835  &    47.973  &    33.706  &    22.033  &    12.953  \\
    Top1(\%)  & 69.07 & 67.256 & 65.78 & 62.3  & 56.036 & 66.404 & 64.364 & 59.928 & 53.112 & 43.002 & 59.502 & 57.452 & 52.608 & 45.316 & 34.88 & 54.43 & 51.214 & 46.59 & 39.328 & 27.2 \\
    \bottomrule
    \end{tabular}%
  \label{tab:desingSpace}%
  \end{adjustbox}
\end{table*}%

\textbf{Classifier CU:} The last Compute Unit is the Classifier CU, which concludes the DSCNN implementation. Fig.~\ref{fig:Net_Classifier} represents the \mn~Classifier CU. Similar to others, this CU is parameterized such that the parallelism across the computing core can be adjusted based on the available hardware resources. Classifier CU comprises compute-bound operations and has a similar configuration to the \pw~convolutional operators.

\subsubsection{Host Code Scheduling and CUs Management}
Finally, the Network Soc Compiler also manages the host-level scheduling of CUs. Fig.~\ref{fig:runtime} visualizes the CUs scheduling and their memory footprints on shared memory. The host or PS initializes the DDR with network models and quantization parameters. \red{The DeepDive back-end generates the memory layout so that the network data region is shared between PL and PS. Therefore at each CU invocation, the PS only passes the data pointer, and the PL fetches the data based on the provided pointer rather than copying the data to its region. This memory layout will remove the necessity of copying data between the PL and PS memory region.} The host starts scheduling procedure by configuring the Head CU with appropriate memory pointer addresses, offsets, network parameters, and network configuration, i.e., $M$, $N$, $H$, which are compiled into network configuration header files. When Head CU completes execution, it writes back the data in feature tensors and interrupts the host CPU. Following the same trend, the host will schedule the Body CUs for $j$ times, where $j$ is the number of Body CU invocations calculated based on CU's mapping. Host CPU then schedules the Tail CU, which executes the compute-bound operations quickly. And finally, the last call is to the Classifier CU, which will update the content of feature tensor needed by the softmax layer to calculate the confidence. Host CPU creates a sequential yet fused scheduling and management of CUs for DSCNNs.

\section{Experimental Results}\label{sec:ExpResults}
We have chosen the Xilinx Zynq UltraScale+ MPSoC ZCU102 evaluation board, which has XCZU9EG chip, to demonstrate the capabilities of DeepDive. The ARM processors host Ubuntu 16.04, running at 1.2GHz; the OS can program the FPGA fabric at runtime. We also use Vivado HLS 2018.3 to synthesize the network models compiled by DeepDive. The FPS and power consumption reported for DeepDive are based on \qnet~accelerator running at 200MHz. We targeted \mn~and \effnet~networks as two cases of DSCNNs. The Top-1 accuracy reported in this section is based on training and evaluating the network on the ImageNet dataset. Since the input image has a square shape, we reported only $H$ as input feature size. Later, we elaborate the design exploration and implementation of each one of these networks as a case study.

\begin{figure}[h]
    \centering
    \subfigure[Top1 Accuracy for three different design points.]{
    \includegraphics[width=.5\textwidth, trim= 1 1 1 1,clip, keepaspectratio]{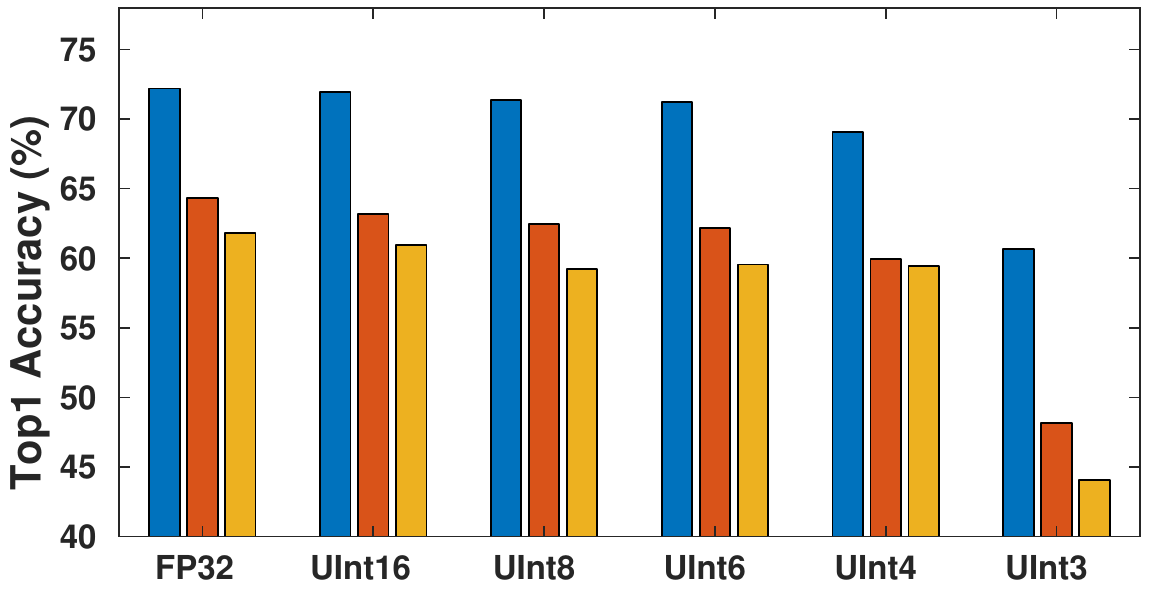}
   \label{fig:top1_int}
    }
    \\
    \subfigure[Model size for three different design points.]{
        \includegraphics[width=.5\textwidth, trim= 1 1 1 1,clip, keepaspectratio]{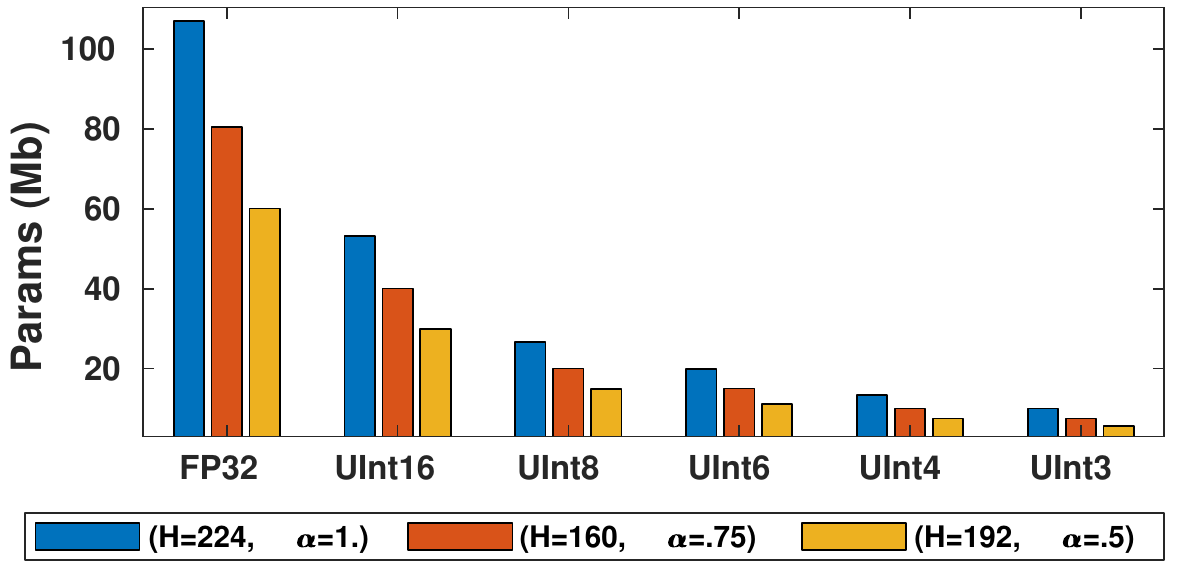}
    \label{fig:model_int}
    }
\caption{\red{The effect of different computation types on Top1-accuracy and model size. Based on Fig. \ref{fig:top1_int}, UInt4 has almost accuracy similar to floating-point, while a notable drop can be observed for UInt3. Also, Fig. \ref{fig:model_int} shows integer quantization causes an exponential decrease in the model size.}}
    \label{fig:diffBits}
\end{figure}

\begin{figure}[h]
    \centering
    \includegraphics[width=.5\textwidth,trim= 1 1 1 1,clip]{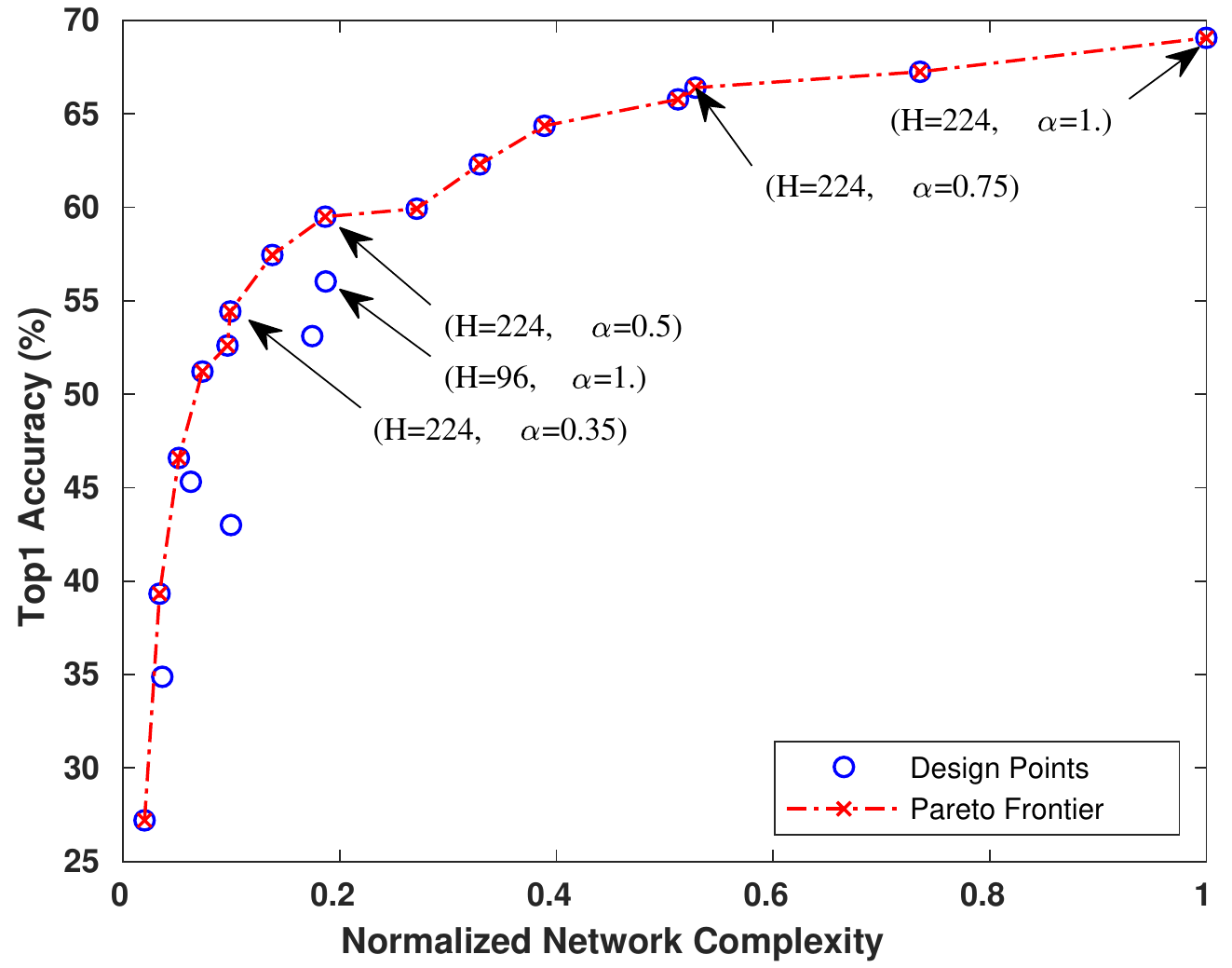}
    \caption{Top1-Network Complexity Pareto front. Design point ($H=96, \alpha=1$) has similar network complexity while is Top1 accuracy is less than ($H=224, \alpha=0.5$).}
    \label{fig:paretoFront}
\end{figure}

\subsection{Case Study: \mn}\label{subSec:CaseStudy}
The procedure starts from a PyTorch model of \mn, pre-trained on ImageNet. At DeepDive's front-end, we configured the FPGA-aware training for different $BW$ based on the channel-wise asymmetric ranged linear quantization. Fig.~\ref{fig:diffBits} shows the Top-1 accuracy for \mn~when its $\alpha=0.75$ and $H=160$. As can be seen, DeepDive maintains accuracy with respect to FP32 by reducing the $BW$ to 8 for first \nc, and 4 for the rest of the layers, respectively. The per layer-specific quantization compresses the model size with a ratio of 8, with 4.4\% degradation in Top1 accuracy. The results demonstrate a dramatic drop in accuracy for $BW=3$. For the rest of this case study, $BW=4$, as it achieves competitive accuracy with considerably smaller model size.

\subsubsection{Design Exploration}
The front-end is configured to re-train, quantize, and calibrate the network for different $\alpha$ and $H$ values. Table~\ref{tab:desingSpace} summarizes the model size, operation numbers and Top1 accuracy per each design point. Based on Table~\ref{tab:desingSpace}, we observe that model size is only effected by $\alpha$, while the number of operation number is a function of both $\alpha$ and $H$. Top1 accuracy is also a function of both $H$ and $\alpha$; however, it is not a linear relationship. For instance, design point ($H=224, \alpha=0.75$) has better Top1 accuracy compared to design point ($H=160, \alpha=1$) while its model size is 33\% less than the latter one. Therefore, we introduce the network complexity as the product of the network model size and network operation number to consider both of them.

\begin{table*}[htbp]
\centering
\caption{Effect of altering $\alpha$ and $H$ for fixed $BW=4$ at 200Mhz on FPS and FPGA Resource Utilization}
\tiny{
\begin{adjustbox}{width=1.0\linewidth,center}
  \begin{tabular}{lccccc|ccccc|ccccc}
    \toprule
    \toprule
    $\alpha$ & \multicolumn{5}{c}{0.75}              & \multicolumn{5}{c}{0.5}               & \multicolumn{5}{c}{0.35} \\
    \midrule
\textit{H} &224 &192 &160 &128 &96 &224 &192 &160 &128 &96 &224 &192 &160 &128 &96 \\
FPS &11 &14 &18 &22 &28 &16 &19 &25 &30 &37 &20 &25 &31 &40 &51 \\
Power(mW) &460 &450 &440 &370 &350 &400 &320 &310 &300 &290 &270 &270 &260 &250 &250 \\
DSP(\%) &57 &57 &58 &57 &57 &37 &37 &37 &37 &37 &24 &24 &24 &24 &24 \\
LUTs(\%) &75 &74 &76 &74 &74 &71 &70 &70 &70 &70 &68 &67 &67 &67 &67 \\
BRAM(\%) &96 &96 &97 &92 &90 &92 &91 &89 &88 &87 &84 &84 &82 &81 &80 \\
    \bottomrule
    \end{tabular}%
  \label{tab:util_power}%
  \end{adjustbox}
  }
\end{table*}%

Fig.~\ref{fig:paretoFront} depicts the Top1-Network Complexity Pareto front. The network complexity helps the front-end to measure the final hardware complexity at a higher level of abstraction. We annotate the starting point of each $\alpha$ in this figure and one non-Pareto point for the sake of comparison. Here, we observed that the design point ($H=96, \alpha=1$) has approximately the same network complexity with respect to ($H=224, \alpha=0.5$), while its Top1 accuracy is almost 4\% less than top achievable accuracy.

\subsubsection{Execution Results and Comparison}
\red{This subsection evaluates DeepDive's execution performance for \mn~on the Hardware Accelerator, different energy-efficient design points implementations, and finally provides a comparison against two other FPGA accelerators \cite{DBLP:vta, mob}. Since there are no other solutions that support both \mn~and \effnet, we also compare it against Nvidia's Jetson Nano as existing state-of-the-art system.}

\begin{figure}[h]
\centering
\includegraphics[width=0.5\linewidth, trim= 20 25 20 20,clip, keepaspectratio]{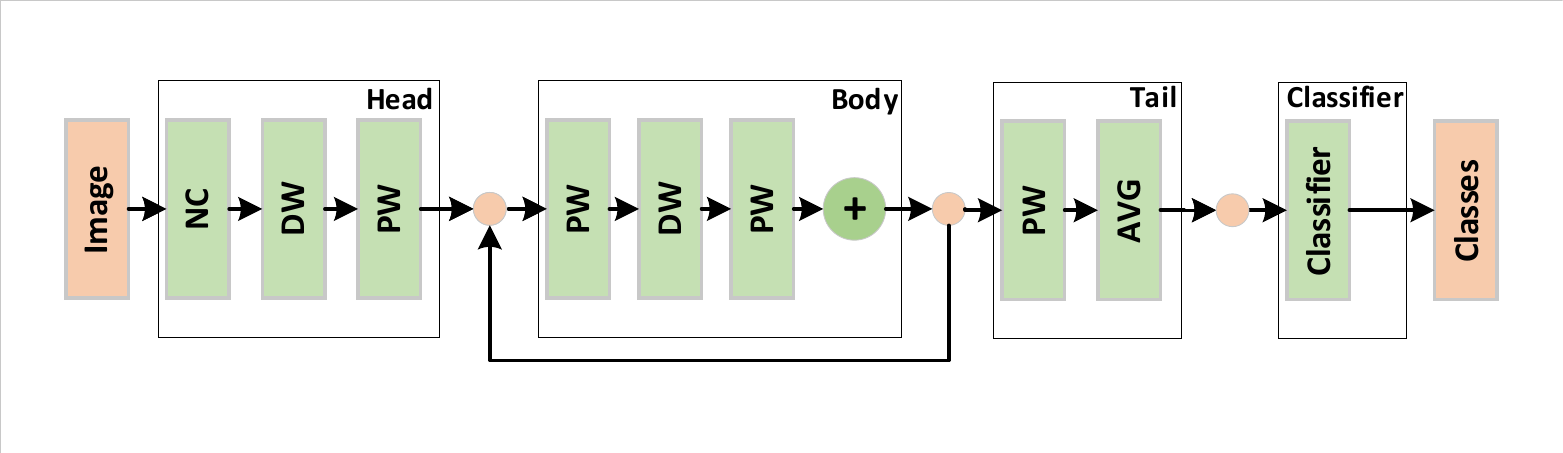}
\caption{\mn~mapped to CUs.}
\label{fig:mnToCUs}%
\end{figure}

\textbf{Mapping:}~As discussed in section~\ref{sec:DD_B}, based on the network graph generated by Network Compiler, DeepDive's back-end identifies the mapping between the convolutional operators and heterogeneous CUs. Fig.~\ref{fig:mnToCUs} reveals the mapping of \mn~to \htr~CUs. The Head, Tail, and Classifier CU are scheduled only once, but the Body is scheduled 16 times. Because of this, DeepDive allocates maximum resources to the Body CU to gain maximum performance. It makes the body CU support both memory-bound and compute-bound operations. For $\alpha=1.0$, DeepDive was not able to fit the design in XCZU9EG SoC chip. If we configure the DeepDive to select different values, less than $N_{max}$ per operator, we observed a significant degradation in the final accelerator performance. Therefore, for the rest of this section, we did not consider these design points.

\textbf{Energy efficiency:}~Here we configure the back-end to compile different network architecture by altering $\alpha$ and $H$. Multiple fully functional execution instances have been created for all configurations in Table~\ref{tab:desingSpace}, except when $\alpha=1.0$.

\begin{figure}[h]
    \centering
    \subfigure[Board power consumption in idle mode]{
        \includegraphics[width=0.3\linewidth,trim= 200 200 100 200,clip, angle =90]{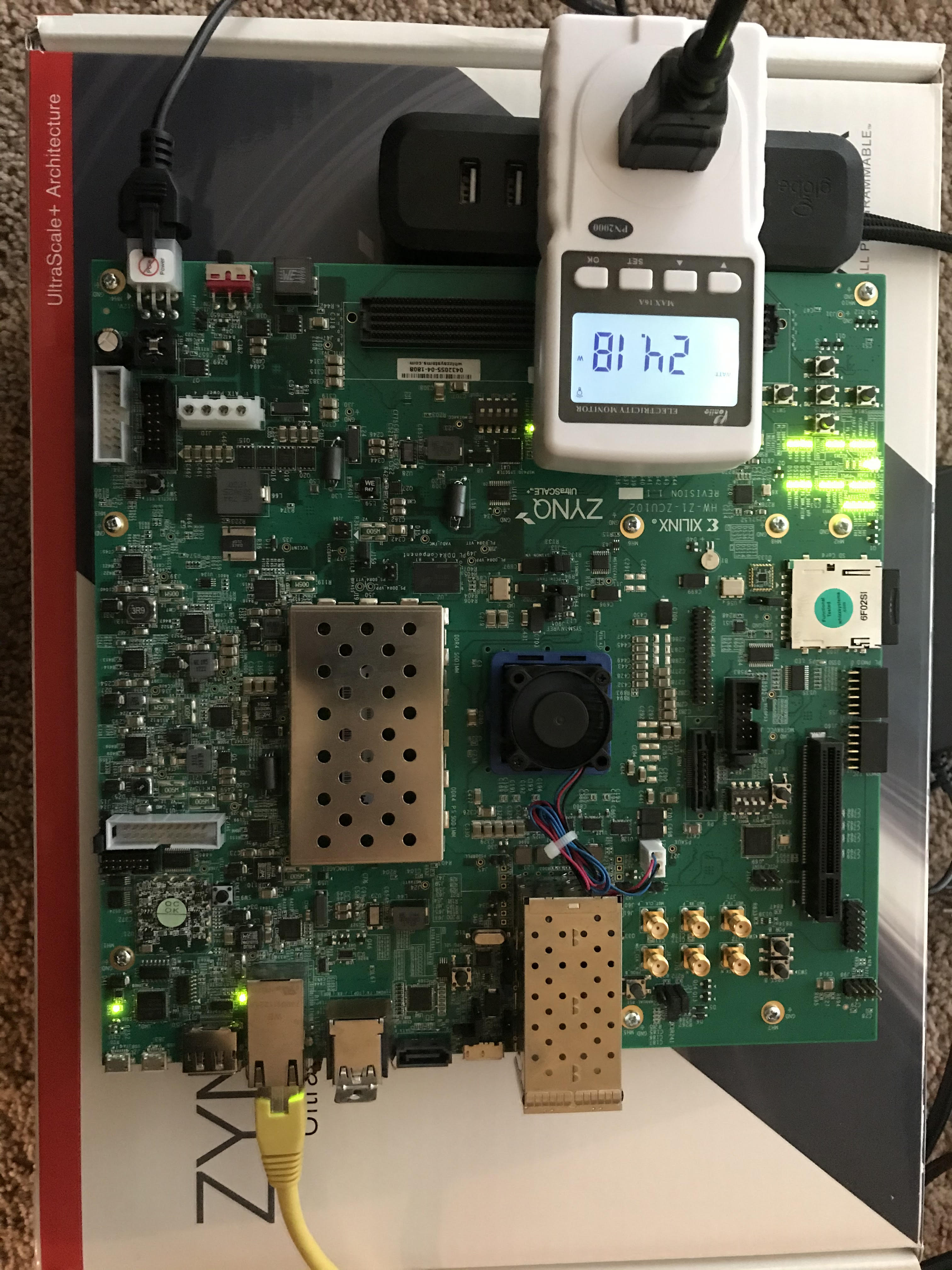}
    \label{fig:pwrIdle}
    }
    \qquad
    \subfigure[Board power consumption for DeepDive inference]{
        \includegraphics[width=0.3\linewidth,trim= 200 200 100 200,clip, angle =90]{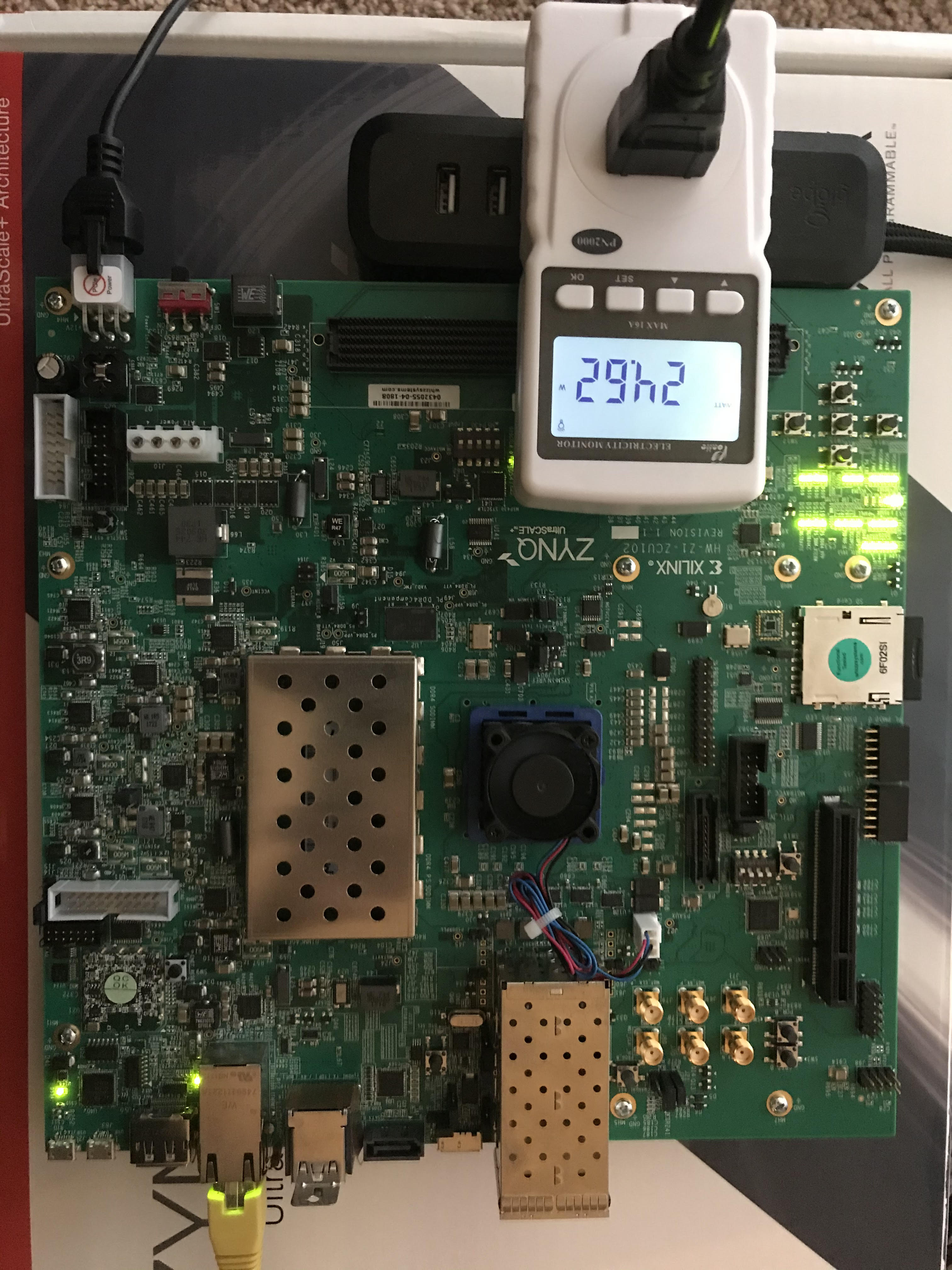}
    \label{fig:pwrInfr}
    }
    \caption{\textbf{Power consumption difference between two mode.}}
    \label{fig:pwr}
\end{figure}

\begin{figure}[h]
    \centering
    \includegraphics[width=.5\textwidth,trim= 0 0 0 0,clip]{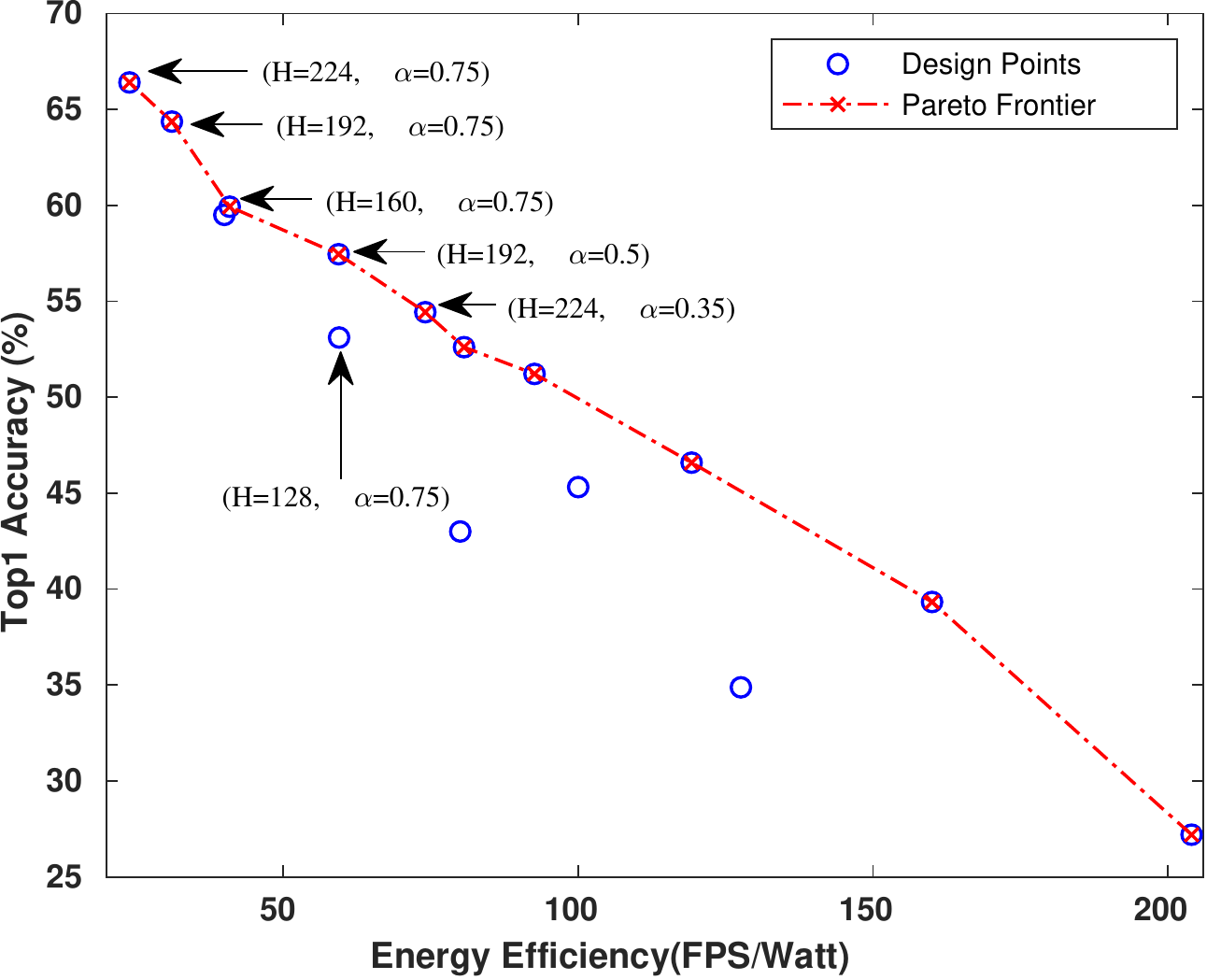}
    \caption{\textbf{Top1-Energy Efficiency Pareto front. Design point ($H=192, \alpha=0.5$) \red{and ($H=128, \alpha=0.75$)} has similar energy efficiency while Top1 accuracy \red{for ($H=192, \alpha=0.5$) is~}more.} }
    \label{fig:pareto_fps}
\end{figure}

Table \ref{tab:util_power} summarizes the power consummation, FPS, and hardware utilization. The power is measured using a power monitoring device, as shown in Fig.~\ref{fig:pwr}. The measured power is the difference of the idle power dissipation of the board and the power consumed by the DeepDive accelerator while running inference. \red{This power is consumed by MPSoC (ARM cores + FPGA fabric), memory hierarchies, and shared DDR memory during the inference.} Resource utilization is directly proportional to $\alpha$, while the power is a function of both $\alpha$ and the input resolution $H$. Design point ($H=96, \alpha=0.35$) has the lowest power consumption at 250mW, as compared to ($H=224, \alpha=0.75$) with the highest power consumption at 460mW. 
Fig.~\ref{fig:pareto_fps} depicts Top1-Energy Efficiency (FPS/Watt) Pareto front. We only annotate the design points that have a higher than 50\% accuracy. DeepDive enables us to understand the relationship between energy efficiency and accuracy. As we can see, design point ($H=160, \alpha=0.75$) has almost same FSP/Watt and Top1 accuracy with ($H=224, \alpha=0.5$). Similarly, the next design point, ($H=192, \alpha=0.5$), can improve energy efficiency by 45.14\%, while the accuracy is dropped by only 2.48\%. Based on the design points provided by DeepDive, it can be observed that by decreasing $\alpha$ and increasing the $H$, we can improve FPS/Watt without sacrificing the Top1-accuracy dramatically. 

\begin{table}[t]
  \centering
  \caption{Power Consumption and delay for \mn }
  \begin{adjustbox}{width=.5\textwidth,center}
    \begin{tabular}{cccc|ccc}
    \toprule
    \toprule
    \multirow{2}[4]{*}{\textit{H}} & \multicolumn{3}{c}{Power(W)} & \multicolumn{3}{c}{Delay(ms)} \\
\cmidrule{2-7}          & Nano(H) & Nano(L) & \multicolumn{1}{c}{DeepDive} & Nano(H) & Nano(L) & DeepDive \\
    \midrule
    224   & 5.49  & 2.64  & 0.46  & 14.91 & 20.73 & 88.49 \\
    192   & 5.22  & 2.51  & 0.45  & 13.61 & 19.96 & 70.32 \\
    160   & 4.78  & 1.88  & 0.44  & 13.07 & 19.6  & 54.45 \\
    128   & 3.35  & 1.56  & 0.37  & 11.24 & 17.19 & 45.51 \\
    64    & 3.25  & 1.32  & 0.35  & 7.89  & 13.91 & 35.71 \\
    \bottomrule
    \end{tabular}%
  \label{tab:jetsonNano}%
  \end{adjustbox}
\end{table}%

\begin{figure}[h]
    \centering
    \includegraphics[width=.5\textwidth, trim= 1 1 1 1,clip, keepaspectratio]{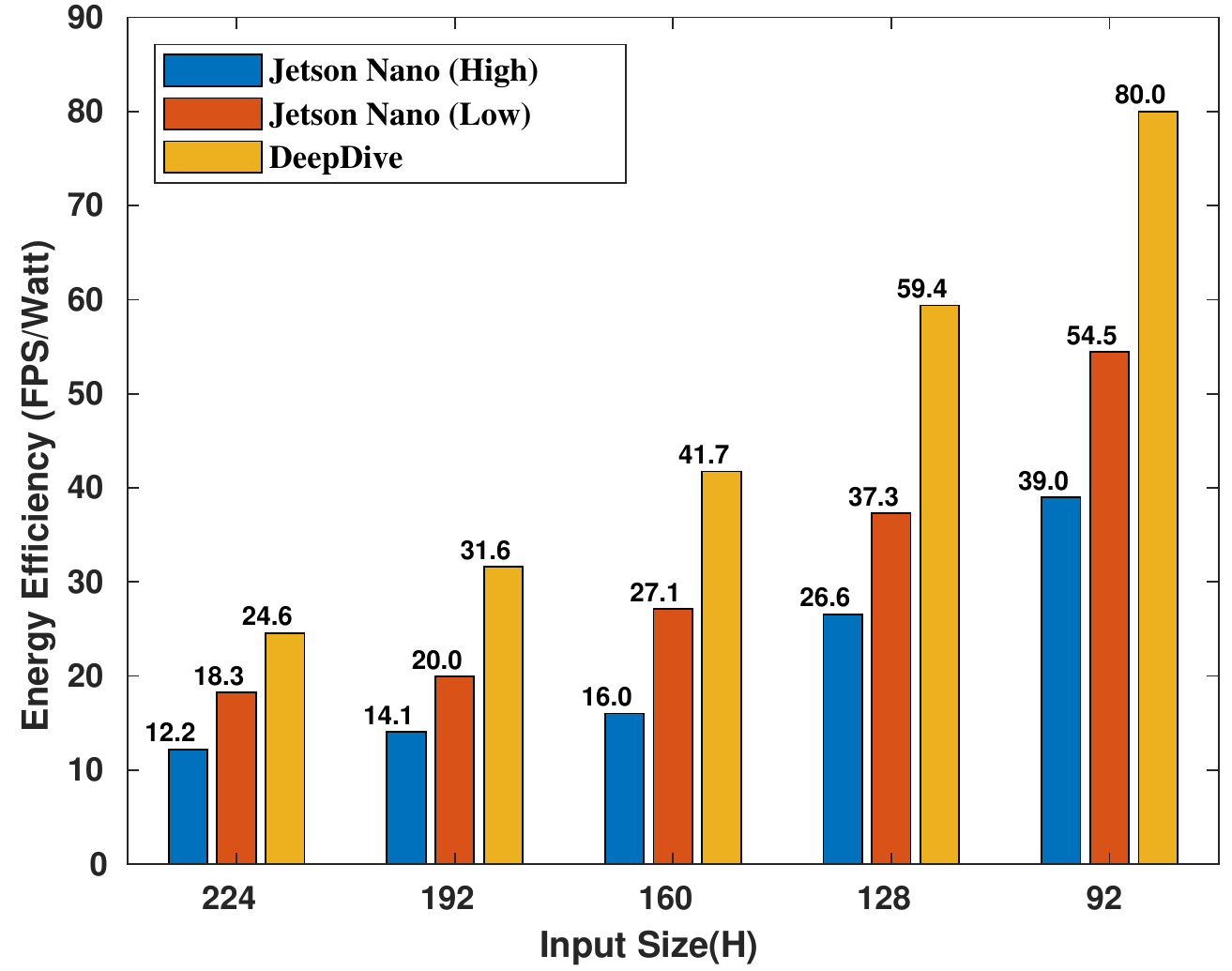}
    \caption{The energy efficiency (FPS/Energy) comparison of DeepDive against Jetson Nano for both high and low power mode.}
    \label{fig:jetsonNano}
\end{figure}

\textbf{Comparison:}~To showcase the energy efficiency of DeepDive, we compare its FPS/Watt against off-the-shelf Nvidia Jetson Nano IoT Edge Device. We mapped the design points of Table \ref{tab:util_power} to TensorRT and obtained the metrics after its graph optimization and quantization. Similar to the DeepDive, we calculate the power consumption only for inference time. We compared the delay and power consumption between DeepDive and Jetson Nano in two different power consumption modes: high power, and low power. It can be seen that DeepDive consumes a lot less power when compared to Jetson Nano, as depicted in Table~\ref{tab:jetsonNano}. Fig.~\ref{fig:jetsonNano} shows the comparison of the Jetson Nano energy efficiency against DeepDive for different input sizes while $\alpha=0.75$. DeepDive, on average, can improve the FPS/Watt 2.2$\times$ and 1.51$\times$ against high and low power mode, respectively. DeepDive outperforms Nano because \textcircled{\raisebox{-0.9pt}{1}} DeepDive performs extreme bit quantization as opposed to nano which uses FP16; \textcircled{\raisebox{-0.9pt}{2}} Although, TensorRT optimized the network model to fuse convolutional operators, DeepDive groups the convolutional operators in heterogeneous CUs at higher granularity. This heterogeneity effectively reduces the shared memory transactions and overlaps both computing and memory latency; \textcircled{\raisebox{-0.9pt}{3}} DeepDive provides a customized dataflow for \dw~separable \cn~as opposed to Jetson Nano which performs general matrix multiplication for \dw~\cn~due to fixed systolic array implementation.


Table~\ref{tab:deepDiveVsOthers} provides a comparison between DeepDive configured with ($H=224, \alpha=0.75)$ design and other similar accelerators. Since VTA's \cite{DBLP:vta} architecture does not support \dw~\cn, they modify the MobileNet to have group convolutions instead of \dw~convolutions, coined MobileNetG. Their MobileNetG was not accessible; hence, there was no chance to present a straightforward comparison. However, we realized that ResNet-18 has almost same inference latency when compared to MobileNetG based on their results, so we decided to compare the energy efficiency of VTA running ResNet-18. As we can see, DeepDive can improve energy efficiency 2.27$\times$. The instruction-based scheduling approach, and versatile systolic array adopted by VTA, both need to consume more power to decode instructions and map layers to the ALU sequentially, which leads to more shared memory transactions and higher power dissipation. \red{Similarly, we compare DeepDive with the hardware accelerator presented by \cite{mob}. DeepDive outperforms \cite{mob} by 37.25$\times$ in energy efficiency. This improvement is because of two main reasons: \textcircled{\raisebox{-0.9pt}{1}} Extreme bit-quantization, BN, and ReLU activation fusion accomplished by front-end which increases the efficiency of the hardware accelerator. \textcircled{\raisebox{-0.9pt}{2}} DeepDive groups the convolutional operators in the CUs at higher granularity to overlap the memory transactions and computations.} 

\begin{table}[htbp]
  \centering
  \caption{\red{Performance Comparison in Classification}}
  \begin{adjustbox}{width=1.0\linewidth,center}
    \begin{tabular}{ccccccc}
    \toprule
    \toprule
    Design & Network & Platform & Freq. (MHz)& Speed (FPS) & Power (W) & Energy Efficiency (FPS/W) \\
    \midrule
    VTA \cite{DBLP:vta} & ResNet-18 & ZCU102 & 200 & 15.44 & 1.47 & 10.51 \\
    \cite{mob} & 0.5 MobileNet & ZYNQ 7Z045 & 100 &1.38 & 2.15 & 0.6418 \\
    Ours & \mn & ZCU102 & 200 & 11 & 0.46 & 23.91 \\
    \bottomrule
    \end{tabular}%
  \label{tab:deepDiveVsOthers}%
  \end{adjustbox}
  \vspace{-10pt}
\end{table}%

\subsubsection{6-bit Data-path}
\red{To showcase the ability to create random bit data-paths, We reconfigure DeepDive to generate and synthesize the \mn~for $BW=6$ to understand the effect of different bit resolution on the final Top1 accuracy and the hardware efficiency. We observe that $BW=6$ can improve the Top1 accuracy by 1.49\%, while the effectiveness of the hardware (FPS/W) drops by 4.88\% on the average.}

\red{Overall, DeepDive improves hardware efficiency by adopting customized functional blocks for \dw~and \pw~convolutions. Heterogeneous CUs also remove unnecessary memory transactions between the PL and shared memory by fused pipeline execution across layers within a block which decreases the power consumption, while improving the overall system performance.}

\subsection{Case Study: EfficientNet}\label{subSec:CaseStudy_effnet}

The baseline \effnet~model was intentionally designed to be larger than \mn. While this might be ideal for state-of-the-art accuracy, it was not suitable for low-power embedded devices. Taking advantage of the compound model scaling factors introduced in \cite{EfficientNet}, we were able to compress the model using smaller $\alpha$, network depth, and $H$, to achieve a model size capable of running on edge devices. The algorithmic details and hardware resource utilization of this model can be seen in Table \ref{tab:effNetStats}. 

\begin{figure}[h]
\centering
\includegraphics[width=.5\textwidth, trim= 25 20 20 20,clip, keepaspectratio]{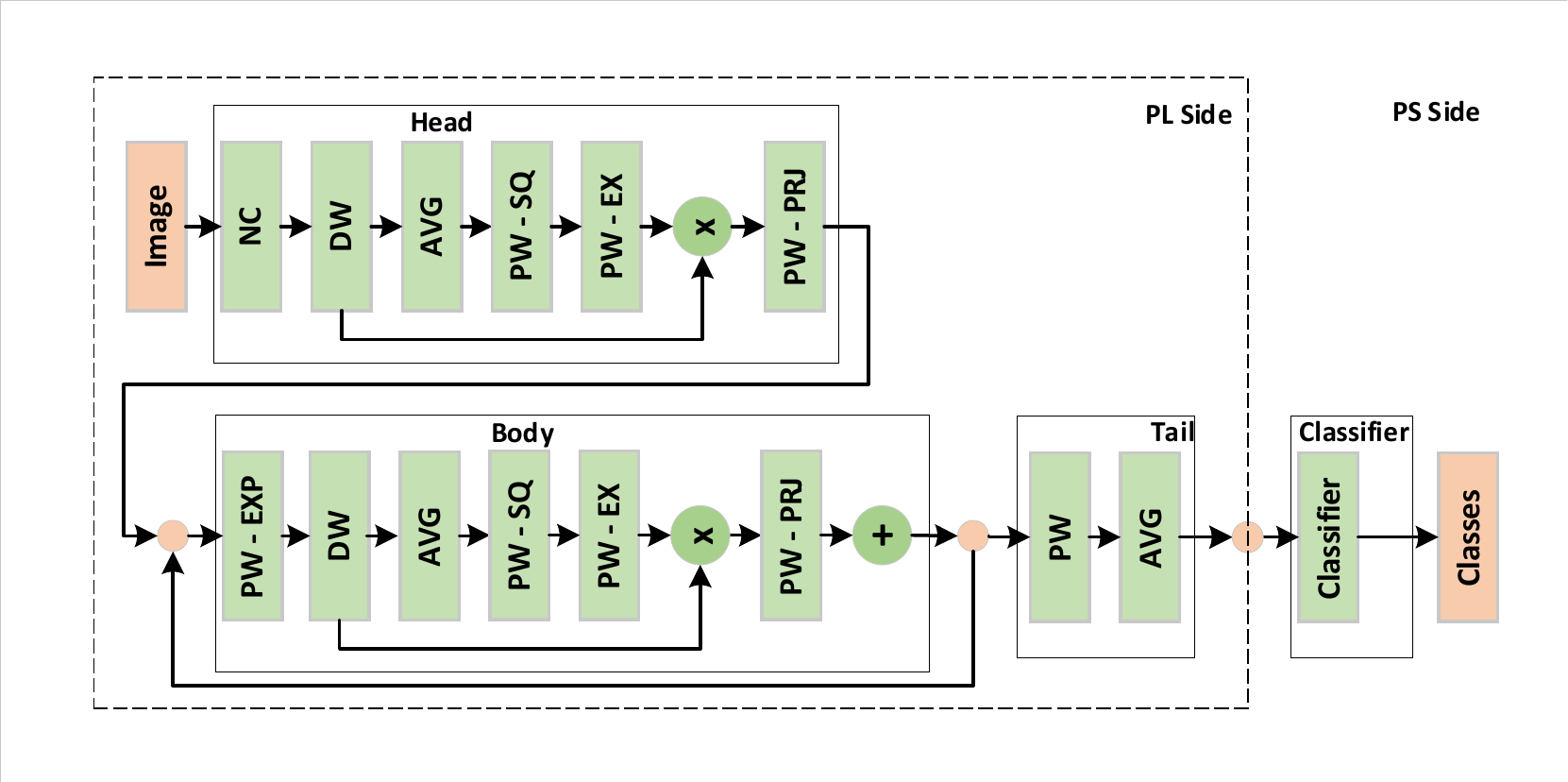}
\vspace{-.30cm}
\caption{\effnet~mapped to CUs.}
\label{fig:efnToCUs}%
\end{figure}


\begin{table*}[h]
  \centering
  \tiny{
  \caption{Compressed \effnet~Algorithmic Specs and FPGA Resource Utilization with fixed BW = 4, Frequency = 200 MHz}
  \begin{adjustbox}{width=1.0\linewidth,center}
    \begin{tabular}{cccc|ccccc}
    \toprule
    \toprule
    \multicolumn{4}{c}{Algorithmic Parameters} & \multicolumn{5}{c}{Hardware Parameters} \\
    \midrule
    \textit{H} & Parameters (Mb) & \#Ops (M) & Top1 (\%) & FPS & Power (mW) & DSP (\%) & LUTs (\%) & BRAM (\%) \\
    \midrule
    128 & 7.81 & 4.914 & 55.02 & 35 & 150 & 90 & 80 & 68 \\
    \bottomrule
    \end{tabular}%
  \label{tab:effNetStats}%
  \end{adjustbox}
  \vspace{-10pt}
  }
\end{table*}%

\textbf{Mapping:}~\effnet~is structurally different as compared to \mn. Fig.~\ref{fig:efnToCUs} shows the mapping of \effnet~to the CUs. The squeeze and excitation convolutional operators are represented as PW-SQ and PW-EX, respectively. DeepDive takes advantage of \effnet~architecture by fusing more convolutional operators together. \effnet~comparatively has a larger body than the \mn, with six layers fused. This mapping helps in achieving better performance by reducing more memory transactions by invoking the Body CU only nine times. For the case of \effnet, we excluded the classifier from mapping and also comparison.


\begin{table}[h]
  \centering
  \caption{Power Consumption and delay for Compressed \effnet}
  \begin{adjustbox}{width=.5\textwidth,center}
    \begin{tabular}{cccc|ccc}
    \toprule
    \toprule
    \multirow{2}[4]{*}{\textit{H}} & \multicolumn{3}{c}{Power(W)} & \multicolumn{3}{c}{Delay(mS)} \\
\cmidrule{2-7}          & Nano(H) & Nano(L) & \multicolumn{1}{c}{DeepDive} & Nano(H) & Nano(L) & DeepDive \\
    \midrule
    128   & 5.61  & 2.22  & 0.15  & 6.581 & 12.6 & 28.57 \\
    \bottomrule
    \end{tabular}%
  \label{tab:dd_effnet_nano}%
  \end{adjustbox}
\end{table}%

\textbf{Energy Efficiency:}
As we can see in Table~\ref{tab:effNetStats}, the number of body CU invocation is 1.78$\times$ less than \mn, which leads to less power consumption and higher FPS due to fewer memory transactions. Table~\ref{tab:effNetStats} shows DeepDive reaches to 35 FPS for a power consummation of 150mW. This model gives us the Energy Efficiency of 233.3 FPS/Watt.

\vspace{+10pt}
\textbf{Comparison:} Table~\ref{tab:dd_effnet_nano} compares the FPS/Watt against Nvidia Jetson Nano. For \effnet, DeepDive can improve the FPS/Watt 8.6$\times$ and 6.7$\times$ against high and low power mode, respectively. Based on the massively fused layers in Body CU, fewer memory transactions translates to more energy-efficient hardware.


\section{Related Work}\label{sec:relatedWork}
Modern CNN accelerators can be divided into two main categories: single compute engine \cite{dnnweaver:micro16, angeleye, dnnbuilder, snowflake, sysarrayaccel, Caffeine}, and multiple streaming compute engines \cite{DeepBurning, FINN, fpgaConvNet,haddoc2, FINN-R, dnnbuilder}. Single compute-engine accelerators are typically a systolic array of processing elements (PEs) that execute the target CNN layer-by-layer sequentially. They have a versatile solution to support different CNNs with the cost of some execution deficiencies. In contrast, streaming architectures consist of multiple dedicated hardware blocks, customized for the target CNN's layers running in producer/consumer fashion. While achieving relatively higher efficiency, they have less scalability to support different networks \cite{stream,stream2}.

Many recent frameworks have proposed a vertical design flow from algorithm to the hardware \cite{dnnweaver:micro16, DeepBurning, dnnbuilder, fpgaConvNet, DBLP:vta}. However, the primary focus is on optimizing classical CNNs with dense operation with regular memory access, such as YOLO and ResNet network family. One notable example of single-engine architecture is DNNWeaver \cite{dnnweaver:micro16}. It offers customizable, hand-optimized RTL templates capable of shrinking or expanding the architecture based on the target CNN workload and target device hardware constraints. The templates support common CNN layer operations such as standard convolution, pooling, and batch normalization.  However, the design-flow is not autonomous as it requires the user to define the network topology and layer structure. Wei et al. \cite{sysarrayaccel} designed a novel 2D systolic array that localizes data shifting to between neighboring PEs. This removes the need for multiplexers and simplifies the routing complexity, allowing for higher throughput. They also employ a custom C-based front-end, which, similar to \cite{dnnweaver:micro16}, requires user interaction to define the nested convolutional loop using custom pragmas in C++. The custom front-end makes it more challenging to integrate with existing high-level DNN libraries (PyTorch, TensorFlow, Caffe, etc). VTA is another recently introduced approach, which presents a versatile hardware solution to support different dense CNNs. VTA enjoys the generality by adapting instruction-based scheduling and flexible systolic array. However, this generality leads to more power dissipation. Another aspect that should be considered is that solutions based on versatile systolic arrays intrinsically do not support \dw~convolutions due to introduced sparsity in these types of convolutions; thus, users need to convert the \dw~convolutions to \gc~to execute a DSCNN on designs similar to VTA. All these succumb to more power dissipation and memory transactions, which lead to having an inefficient hardware solution for DSCNNs.

\red{The design proposed in \cite{autoOpt} presents a framework to minimize the complexity and the model size of dense CNN by mapping normal \cn~to \dw~separable \cn. Similarly, TuRF \cite{noPow} replaces standard \cn~layers with \dw~separable \cn~and applies layer fusion to enhance the performance of dense networks. The design presented by \cite{cnnAccF} is another hardware accelerator based on matrix multiplication and customized adder-tree to support \mn. However, their fixed design platform is not scalable to support fast-growing and forthcoming DSCNNs. A parallel acceleration scheme proposed in \cite{mob}, demonstrates computing reusability with design reconfigurability. However, the accelerator suffers from massive data movements due to frequent reads and writebacks to the DDR because of the lack of fused layer execution. Moreover, the design-flow is not autonomous and requires the user to define the layer structure. A \mn~based hardware accelerator on FP32 computation is presented in \cite{mdpiCnn}. DPU \cite{dpu} is another solution to support \mn~based on an optimized RTL hardware model with a dedicated operator for \dw; however, it cannot be considered as a versatile solution to support DSCNNs due to lake of support for swish activation function and \pw~multiplication. To the best of our knowledge, none of the above approaches present a fully vertical framework to implement the-state-of-the-art DSCNN architectures, e.g., \effnet~family.}
\section{Conclusion}\label{sec:conclusion}

This paper introduced DeepDive, as a fully functional framework for an agile, power-efficient execution of DSCNNs on edge FPGAs. DeepDive offers a vertical algorithm/architecture optimization, starting from the network description model down to full system synthesis and implementation. At the front-end, DeepDive performs  high-level optimization such as BN fusing, and Online channel-wise low-Bit quantization at extremely low-bit resolutions to bring FPGA-awareness when training DSCNNs. At the back-end, Network SoC Compiler receives the design properties from DeepDive's front-end and generates a full design of the system for both hardware model and software host codes. To generate the optimized hardware for DSCNNs, the Network SoC Compiler uses pre-designed micro-architectural blocks for \dw, \pw, and normal \cn~operators. For the results, we have synthesized, executed, and validated two state-of-the-art DSCNNs, \mn~and \effnet~on Xilinx's ZCU102 FPGA board. The execution results demonstrated 47.4 and 233.3 FPS/Watt for \mn~and a compact version of \effnet, respectively. These comparisons showcased how DeepDive improved FPS/Watt by 2.2$\times$ and 1.51$\times$ over Jetson Nano high and low power modes, respectively. It also enhances FPS/Watt about 2.27$\times$ and 37.25$\times$ over two other FPGA implementations.

\red{As future work, we plan to improve the back-end of DeepDive to support cloud-based FPGAs such as Alveo family. We plan to extend support for multiple instances of Body CU to improve both latency and throughput. Each body could have a different level of parallelization based on the knobs introduced in Section \ref{subSec:Conv_op}. The host would also map the IRB layers to the body CUs based on the required computation power. Various Body CUs with varying degrees of parallelization could improve DeepDive without power and hardware resource compromises.}


\bibliographystyle{IEEEtranS}
\bibliography{bibo/bibo}

\end{document}